\documentclass[twocolumn]{aastex6}
\usepackage{amsmath}

\usepackage{color}

\shorttitle{\textit{K2} Observations of an Eclipsing SU UMa Star}
\shortauthors{Littlefield et al.}

\newcommand{\mls}{J0359}

\begin{document}

\title{A Comprehensive \textit{K2} and Ground-Based Study of CRTS J035905.9+175034, \\ an Eclipsing SU UMa System with a Large Mass Ratio}
\author{Colin Littlefield,\altaffilmark{1} Peter Garnavich,\altaffilmark{1} Mark Kennedy,\altaffilmark{2} Paula Szkody,\altaffilmark{3}
Zhibin Dai\altaffilmark{4,5,6} }

\altaffiltext{1}{Department of Physics, University of Notre Dame, Notre Dame, IN 46556}
\altaffiltext{2}{Jodrell Bank Centre for Astrophysics, School of Physics and Astronomy, The University of Manchester, M13 9PL, UK}
\altaffiltext{3}{Department of Astronomy,University of Washington, Seattle, WA 98195}
\altaffiltext{4}{Yunnan Observatories, Chinese Academy of Sciences, 396 Yangfangwang, Guandu District, Kunming, 650216, P. R. China}
\altaffiltext{5}{Key Laboratory for the Structure and Evolution of Celestial Objects, Chinese Academy of Sciences, 396 Yangfangwang, Guandu District, Kunming, 650216, P. R. China}
\altaffiltext{6}{Center for Astronomical Mega-Science, Chinese Academy of Sciences, 20A Datun Road, Chaoyang District, Beijing, 100012, P. R. China}

\begin{abstract}

CRTS J035905.9+175034 is the first eclipsing SU~UMa system for which a superoutburst has been observed by \textit{Kepler} in the short-cadence mode. The light curve contains one superoutburst, eight normal outbursts (including a precursor to the superoutburst), and several minioutbursts that are present before---but not after---the superoutburst. The superoutburst began with a precursor normal outburst, and shortly after the peak of the precursor, the system developed large-amplitude superhumps that achieved their maximum amplitude after just three superhump cycles. The period excess of the initial superhump period relative to the orbital period implies a mass ratio of 0.281$\pm0.015$, placing it marginally above most theoretical predictions of the highest-possible mass ratio for superhump formation. In addition, our analysis of the variations in eclipse width and depth, as well as the hot spot amplitudes, generally provides substantiation of the thermal-tidal instability model. The \textit{K2} data, in conjunction with our ground-based time-resolved spectroscopy and photometry from 2014-2016, allows us to determine many of the fundamental parameters of this system.

\end{abstract}


\section{Introduction}
\label{introduction}

Dwarf novae consist of a low-mass donor star that overfills its Roche lobe and loses mass to a white dwarf primary (WD). An accretion disk forms around the WD and undergoes occasional photometric outbursts, usually with amplitudes of several magnitudes and lasting for several days. One subset of dwarf novae, the SU~UMa systems, show two discrete types of outbursts: normal outbursts and superoutbursts. Superoutbursts are brighter than normal outbursts by about one magnitude and last significantly longer. The light curve of a superoutburst exhibits superhumps, which are periodic modulations whose period is several percent longer than the orbital period. The interval between consecutive superoutbursts is known as a supercycle.

Normal outbursts are postulated to occur as a result of a thermal disk instability \citep{osaki74}. Mass transfer from the secondary causes the disk density to exceed a critical value, leading to the ionization of hydrogen and an ensuing thermal runaway as the disk becomes optically thick. The increased disk viscosity boosts the accretion rate onto the WD before a cooling front extinguishes the outburst. Despite the elevated accretion rate, the disk still gains mass during a normal outburst.

The thermal-tidal instability (TTI) model \citep{osaki89} is the prevailing theory regarding the mechanism of superoutbursts. According to the TTI model, superoutbursts occur when the outer radius of the disk expands to the 3:1 Lindblad resonance, at which point tidal interactions with the donor cause the disk to become eccentric and to undergo apsidal precession. The superhumps appear once the disk becomes eccentric, and the enhanced tidal dissipation of the disk's angular momentum causes increased accretion of the disk onto the WD. Critically, the TTI model makes a fundamental prediction that during any given supercycle, each normal outburst will force the outer disk radius to expand until the 3:1 resonance radius is reached.

During a superoutburst, the superhump period changes, and an O$-$C diagram of the superhump maxima will generally display three distinct regimes: Stages A, B, and C \citep{kato09}. Stage A superhumps are the first superhumps to appear in a superoutburst, and they have the longest period. \citet{ko13} argued that they are observed when the disk eccentricity is confined to the 3:1 resonance and that their period is equivalent to the dynamical precession rate at that resonance. At the end of Stage A, the superhump period decreases abruptly, marking the transition to Stage B. Whereas the superhump period during Stage A is constant, Stage B superhumps usually show a positive period derivative, related to a pressure effect within the disk \citep{ko13}. Finally, Stage C superhumps appear after Stage B, have a shorter period than Stage B, and do not show a period derivative.

The continuous photometry made possible by the \textit{Kepler} satellite has provided significant insight into the behavior of SU~UMa systems, enabling a test of the TTI model's predictions. For example, \citet{ok14} found that each superoutburst in \textit{Kepler} data of V1504~Cyg began with a precursor normal outburst, with superhumps appearing at the maximum of the precursor. They also reported evidence that the system's disk radius increases during its supercycle. Both of these observations supported key predictions of the TTI model.

We report short-cadence \textit{Kepler} observations and ground-based spectroscopy and photometry of the poorly studied cataclysmic variable CRTS J035905.9+175034 (= MLS130302:035906+175034; hereinafter \mls). Our data reveal it to be an eclipsing SU~UMa system with an orbital period of 1.91~h. This object appears in the Sloan Digital Sky Survey Data Release 9 as SDSS J035905.91+175034.47 with $g$=18.50 and colors $u-g$ = 0.91 and $g-r$ = 0.46, and it is listed as a newly discovered cataclysmic variable in \citet{Drake14}.


\section{Data}

\subsection{K2 observations}

The \textit{Kepler} spacecraft observed \mls\ from February 2 to April 24, 2015, as part of Campaign 4 of the \textit{K2} mission \citep{2014PASP..126..398H}. The data were taken in short-cadence mode, with a typical cadence of 58.8~sec per image. The dataset spans 70.9~days, providing nearly continuous coverage during that time. A light curve was constructed by using \textsc{Pyke} \citep{still12} to extract events from the target pixel file. There were occasional, brief gaps in the light curve when onboard thrusters were fired to keep the spacecraft pointing at the campaign field, but these data were removed by deleting observations that had a \textsc{Quality} flag $>0$.


\subsection{Ground-based Spectroscopy}

We obtained spectra of \mls\ with the Large Binocular Telescope and the Multi-Object Dual Spectrograph (MODS) on two occasions. Observations were taken on 2015 October 14 (UT) using MODS1 (SX mirror) through a 1.2-arcsec-wide slit in the grating mode, providing a resolution of 1300. Ten 200-sec exposures were obtained with 90~s of overhead between spectra. The data spanned 40\%\ of the orbital period, including an eclipse.

The LBT also obtained spectra on 2016 January 3 (UT) with the MODS1 spectrograph. A 1.0-arcsec slit was employed with the grating mode, providing a resolution of 1500. On this visit, 27 spectra were obtained, each with a 200-sec exposure time. The CCD was binned by two in the spatial direction, reducing the overhead to 60~s, and the spectral sequence covers slightly more that one full orbit of the system. In both visits, the position angle of the slit was rotated to match the parallactic angle.

\begin{figure}
\includegraphics[width=0.5\textwidth]{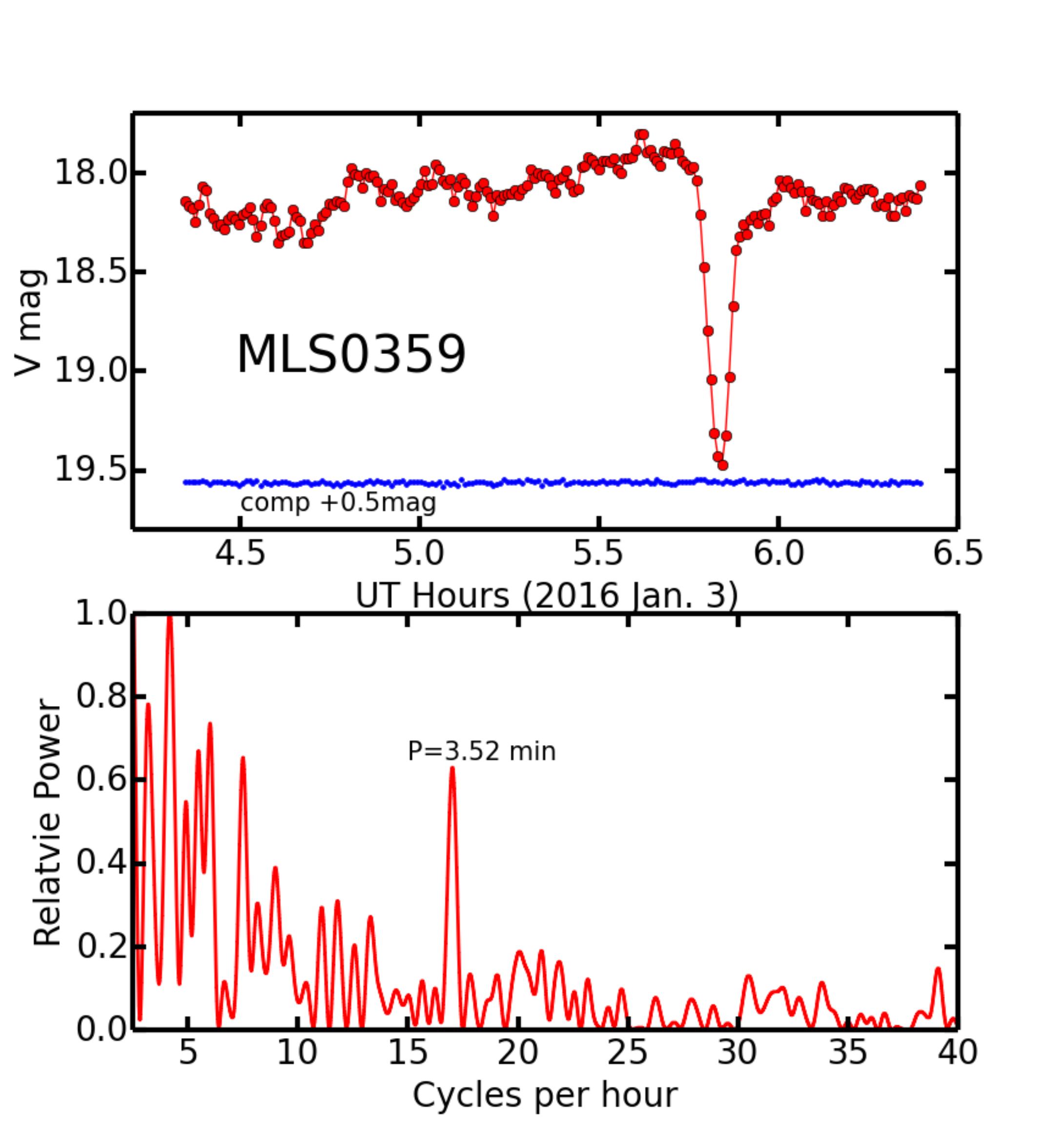}
\caption{{\bf Top:} The light curve of \mls\ obtained with the LBT/LBC simultaneously with the MODS spectroscopy in 2016. The data has twice the cadence of the K2 photometry. {\bf Bottom:} A periodogram of the LBC light curve, showing significant power at 3.5 minutes. The variation is not seen in the \textit{K2} data or the other photometry, suggesting that this is a quasi-periodic oscillation. \label{lbc_LC}}
\end{figure}

\floattable
\begin{deluxetable}{lcl}
\label{obs_log}
\tablewidth{0.5\textwidth}
\tablecaption{Summary of Observations}
\tablehead{
\colhead{UT Date} & \colhead{Site} & \colhead{Observations}} 
\startdata
Sep 25, 2005 - Dec 23, 2013 & CSS+MLS & sporadic photometry \\
Aug 23, 2014 & KPNO 4m & 2 x 1050 s spectra \\
Nov 21, 2014 & APO 3.5m & 18 x 600s spectra \\
Feb 7 - Apr 24, 2015 & K2 & SC continuous photometry \\
Oct 14, 2015 & LBT 8.4m & 10 x 200s spectra \\
Oct 17, 2015 & APO 3.5m & 5 x 600s spectra \\
Dec 18, 2015 & APO 3.5m & 2 x 600s spectra \\
Dec 27, 2015 & China 1m & 207 x 25s photometry \\
Jan 3, 2016 & LBT 8.4m & 27 x 200s spectra + 204 x 15s phot \\
\enddata
\end{deluxetable}

The spectra were bias-subtracted, flat-fielded, extracted using IRAF {\it twodspec} routines, and wavelength-calibrated using Ne and Ar emission arcs taken during the day. We used airglow lines extracted in the sky subtraction to refine the wavelength solution. Finally, the spectra were flux-calibrated from observations of the spectrophotometric standard G191-B2B.

Spectra were also obtained with the Apache Point Observatory 3.5m telescope on 3 occasions during 2014-2015 using the Double Imaging Spectrograph with the high resolution gratings, giving a resolution of 0.6~\AA\ from 4000-5000~\AA\ in the blue and 6000-7200~\AA\ in the red. Two spectra were also obtained with the Kitt Peak 4m telescope and RC Spectrograph using the second order of grating KPC-22b, resulting in a wavelength coverage of 3800-4900~\AA\ at a resolution of 0.7~\AA. As with the LBT data, the spectra and calibration lamps and flux standards were reduced using IRAF routines. The observations are summarized in Table~\ref{obs_log}. Since the spectra were similar to the LBT but with reduced SNR and time resolution, most of the analysis in this paper uses the LBT spectra.

\subsection{Ground-based Photometry}

We acquired photometry on 2015 Dec. 27 using a 1m Cassegrain telescope and an Andor DZ936 camera at Weihai Observatory at Shandong University in Weihei City, China \citep{hu}. The exposure time was 25~s, and the observations lasted for 1.5~h. One eclipse was observed, with the light curve showing no evidence of a pre-eclipse hump.

During the 2016 LBT run, the Large Binocular Camera (LBC) obtained photometry simultaneously with the spectra. The exposure time was 15.24~s through a Bessel $V$ filter, and to improve the time resolution, only 1000 rows of the central chip were read out. In all, we took 204 images between 4:20 and 6:24 UT, with an average time between exposures of 35.6~s. We performed aperture photometry of \mls\ and nearby stars, calibrating the data to the $V$-band using the APASS catalog. The typical brightness of \mls\ outside of eclipse ranged between $V$ magnitudes 17.9-18.2. The light curve included one eclipse and a low-amplitude pre-eclipse hump attributable to the stream-disk hot spot. After removing observations obtained during eclipse and detrending the light curve by subtracting a third-order polynomial, we identified a low-amplitude quasi-periodic oscillation with a period of 3.5~min of unknown origin (Fig.~\ref{lbc_LC}).

\begin{figure}
\epsscale{1.2}
\plotone{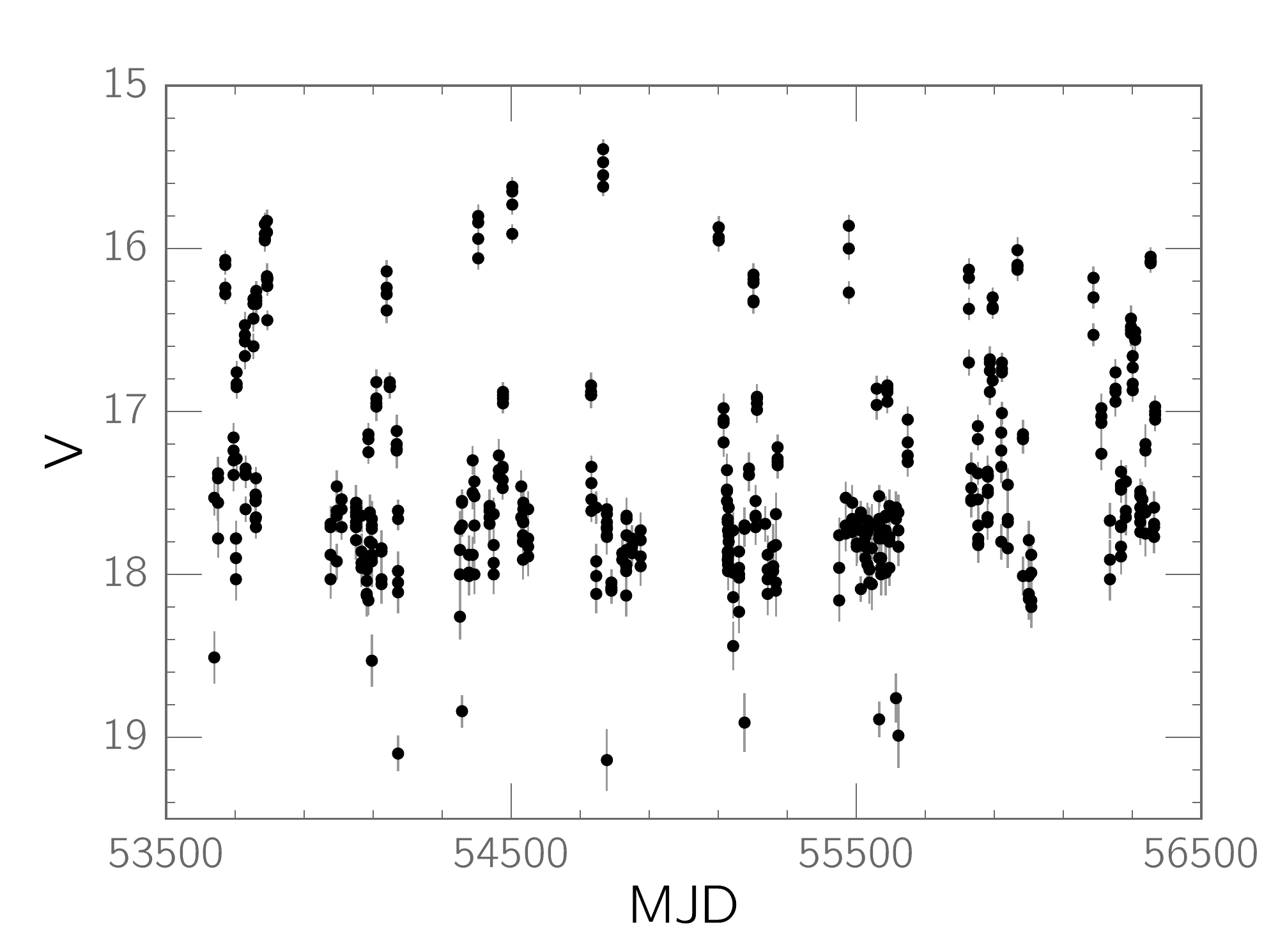}
\caption{The CRTS light curve of \mls. Frequent outbursts are apparent, as are several eclipses. \label{crts}}
\end{figure}

The Catalina Real Time Transient Survey \citep[CRTS;][]{drake09} provides a much longer baseline of observations than our time-series photometry. CRTS observed \mls\ from September 25, 2005 until December 23, 2013 using the Catalina 0.7-m Schmidt telescope and the Mt. Lemmon 1.5-m telescope. These data are shown in Figure~\ref{crts}. Many outbursts are evident in this dataset, along with numerous points obtained during eclipses.

\section{\textit{K2} photometry analysis}

\begin{figure*}[ht]
\includegraphics[width=\textwidth]{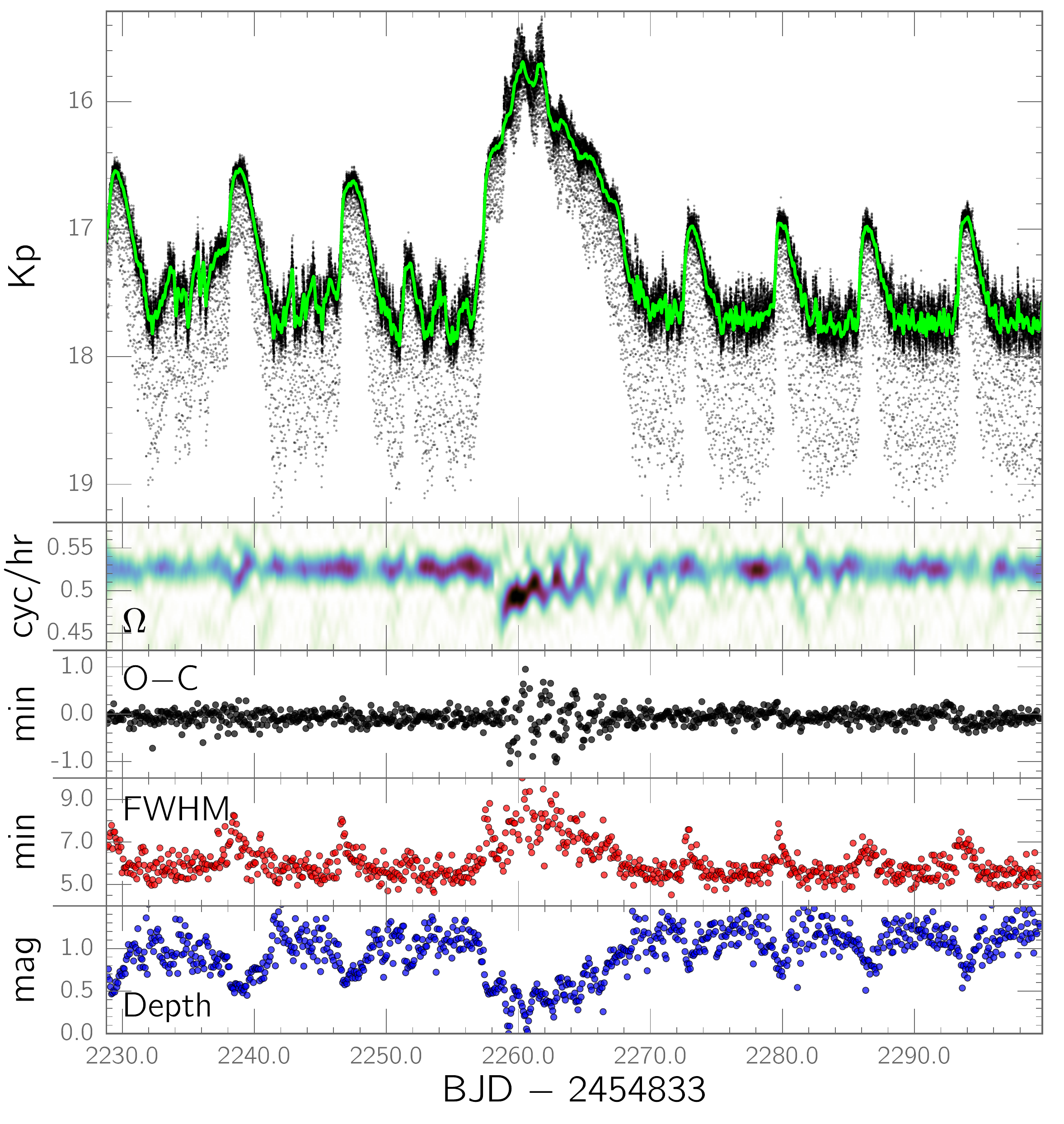}
\caption{The full \textit{K2} light curve of \mls. The smoothed light curve is superimposed as a green line. Minioutbursts give rise to the choppy appearance of the three quiescent segments prior to the superoutburst. From top to bottom, the remaining four panels show the trailed power spectrum around the orbital period's fundamental harmonic ($\Omega$), the eclipse O$-$C, the eclipse full width at half-minimum, and the eclipse depth. The normal outbursts shift the mid-eclipse times to earlier phases because the increased disk luminosity moves the centroid of disk emission away from the hotspot and toward the disk center.\label{full_LC}}
\end{figure*}

As shown in Fig.~\ref{full_LC}, the \textit{K2} light curve contains one superoutburst and eight normal outbursts, one of which is a precursor to the superoutburst. Prior to the superoutburst, there were a number of minioutbursts that recurred quasi-periodically every $\sim$2~d with amplitudes of $\sim$0.5~mag. The minioutbursts had irregular morphologies and partially overlapped with each other to such an extent that it is almost impossible to identify the quiescent level between the first four normal outbursts. There are none after the superoutburst. We  discuss these minioutbursts in Sec.~\ref{mini}.

\begin{figure*}[ht]
\includegraphics[width=\textwidth]{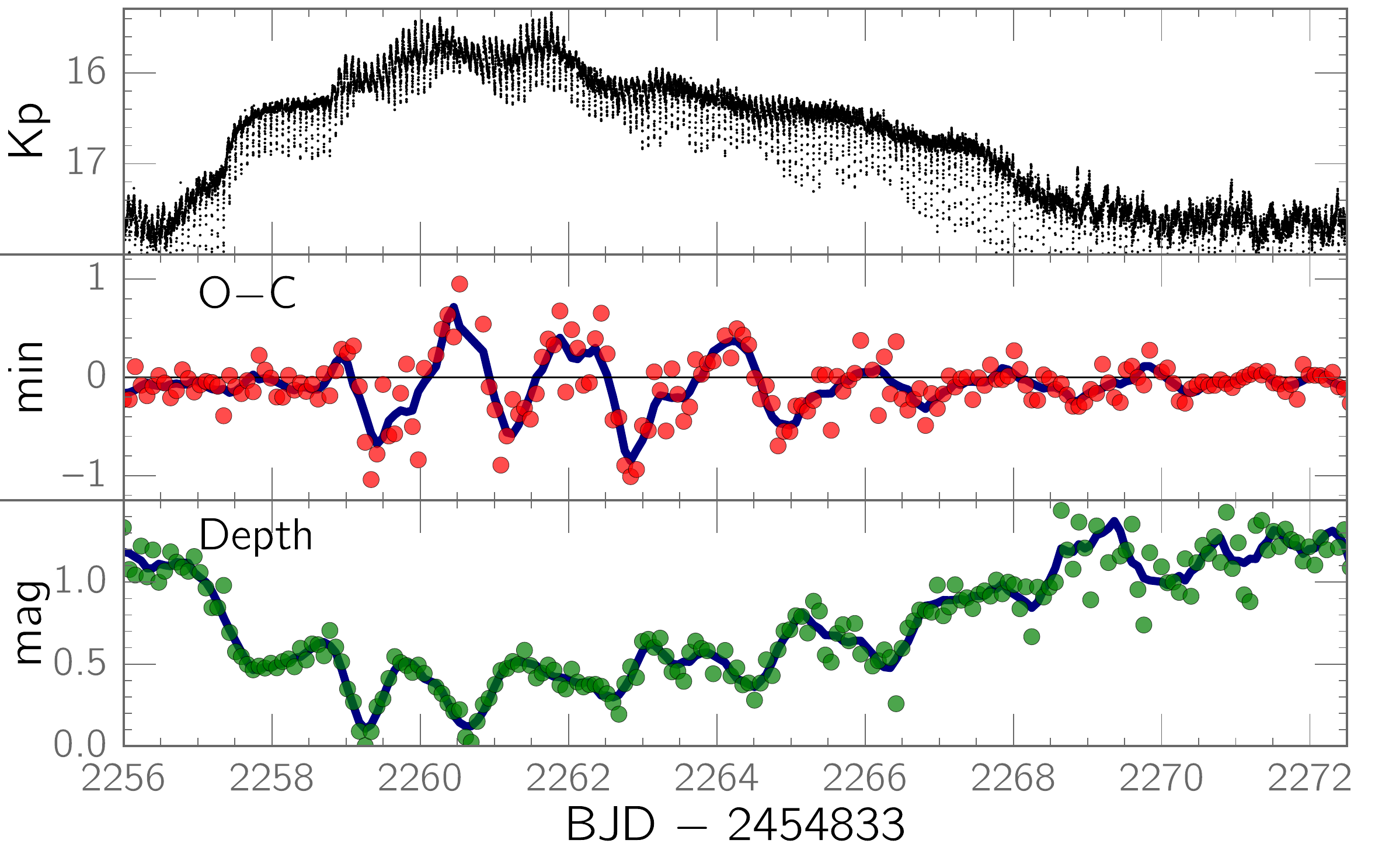}
\caption{Behavior of the eclipse O$-$C timings and depth during the superoutburst. These data were also shown in Fig.~\ref{full_LC}, but we enlarge them here so that details may be more readily discerned. \label{superoutburst_o-c}}
\end{figure*}

To extract the depth, width, and time of minimum light of the eclipses, we fitted polynomials to all eclipses, except for those with incomplete coverage. We visually inspected the fits to ensure their adequacy. The eclipse depth was defined as the ratio of the flux at minimum light to the median out-of-eclipse flux within one orbital cycle of the eclipse; this approach works well for quiescent data and normal outbursts, but it struggles to accurately measure depth when superhumps are present. The width was estimated by numerically determining the full width at half minimum (FWHM) for each polynomial. These parameters are included in the lower panels in Fig.~\ref{full_LC}. Fig.~\ref{superoutburst_o-c} enlarges the superoutburst light curve and the eclipse O$-$C and depth plots so that details may be seen more distinctly.

The times of mid-eclipse are well described by the orbital ephemeris of $$ T_{min}[BJD] =  2457069.9825(2) + 0.079555141(15) \times E, $$ where the numbers in parentheses give the $1\sigma$ uncertainties on the final digits of the corresponding parameters. We show the adequacy of this ephemeris by plotting the eclipse O$-$C residuals in Fig.~\ref{full_LC}.

The trailed Lomb-Scargle periodogram in Fig.~\ref{full_LC} shows the evolution of the power spectrum near the orbital frequency throughout the \textit{K2} light curve. To generate it, we created a smoothed light curve using the LOWESS algorithm \citep{cleveland} and subtracted it from the unsmoothed light curve. We excluded observations between $0.9 < \phi_{orb} < 1.1$ to reduce the signal from eclipses. We then calculated the power spectrum with the brightness expressed in flux (not magnitudes) using a window width of 1.5~d and a step size of 0.15~d. 

Outside of the superoutburst, the power was concentrated at the orbital period and its harmonics. Near the peak of the second normal outburst, the power briefly shifted to a slightly lower frequency for about a half-day before returning to the orbital frequency. At the start of the superoutburst, the superhump frequency appeared, quickly increased in frequency, and stabilized, disappearing as the system returned to quiescence. Throughout the superoutburst, the interaction of eclipses and superhumps caused the power to oscillate between the orbital and superhump frequencies, a phenomenon that persisted until the very end of the superoutburst. Other than the orbital and superhump harmonics, the power spectrum did not show evidence of additional periodicities, such as the candidate quasi-periodic observation detected in one ground-based light curve (Fig.~\ref{lbc_LC}).

\begin{figure}
\epsscale{1.2}
\plotone{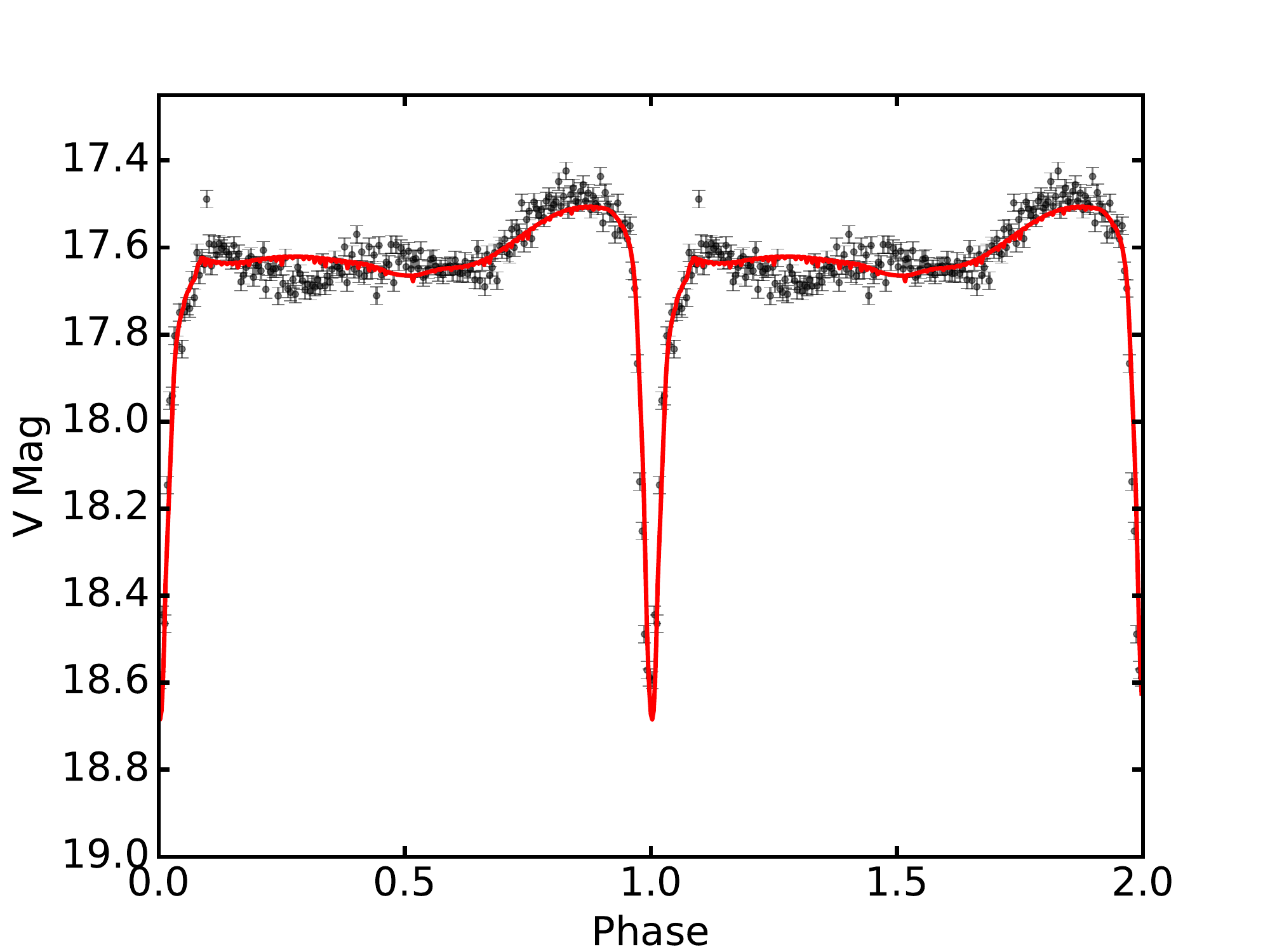}
\caption{A parameterized model (red line) of the quiescent light curve using the \citet{ELC_code} code, yielding a firm constraint of $i = 77^{\circ}\pm2^{\circ}$. \label{ELC_model}}
\end{figure}

The phased light curve of the quiescent orbital modulation (Fig.~\ref{ELC_model}), which showed a deep, 1.3-mag eclipse preceded by an orbital hump from the stream-disk collision, is typical of high-inclination CVs. However, individual orbital cycles showed highly variable morphologies. Dips occurred sporadically at nearly every orbital phase, and the amplitude of the orbital hump was highly variable in individual orbital cycles, ranging from less than 0.1~mag to 0.3~mag. Moreover, the hotspot hump usually peaked before the eclipse, but in a minority of orbits, it appeared to peak at the time of eclipse and was equally visible before and after the eclipse.

We used the Eclipsing Light Curve (ELC) code from \citet{ELC_code} to model part of the quiescent light curve in order to derive physical parameters of the system (Fig.~\ref{ELC_model}). We set the mass ratio to 0.28 (per Sec.~\ref{mass_ratio_sec}), the effective secondary temperature to 3,000 K (in line with the effective secondary temperature for the given orbital period predicted by \citealt{2011ApJS..194...28K}), and X-ray luminosity of the WD to $10^{29}$ erg cm$^{-2}$ s$^{-1}$ (in line with the system being a dwarf nova, which typically have low X-ray luminosities). Based on the lack of WD absorption lines (Sec.~\ref{spectra_sec}), the spectrum suggests that the accretion disk is dominant at optical wavelengths. In keeping with this, we neglected optical light from the WD in our models by setting the effective temperature of the WD to a negative value. After setting the orbital separation to 0.73$R_{\odot}$ and allowing the outer accretion disk radius, inner accretion disk temperature, orbital inclination, and hot-spot parameters to vary, we searched for the best-fit parameters using a Monte Carlo Markov Chain. While many of the model parameters were unconstrained, the inclination was tightly constrained to $i = 77^{\circ} \pm 2^{\circ}.$ The uncertainty was found by measuring changes in the $\chi^{2}$ value based on the number of free parameters, but given that so many other parameters are unconstrained, it is possible that this nominal uncertainty might be underestimated.

\subsection{Normal outbursts}

\begin{figure*}
\epsscale{1.2}
\plotone{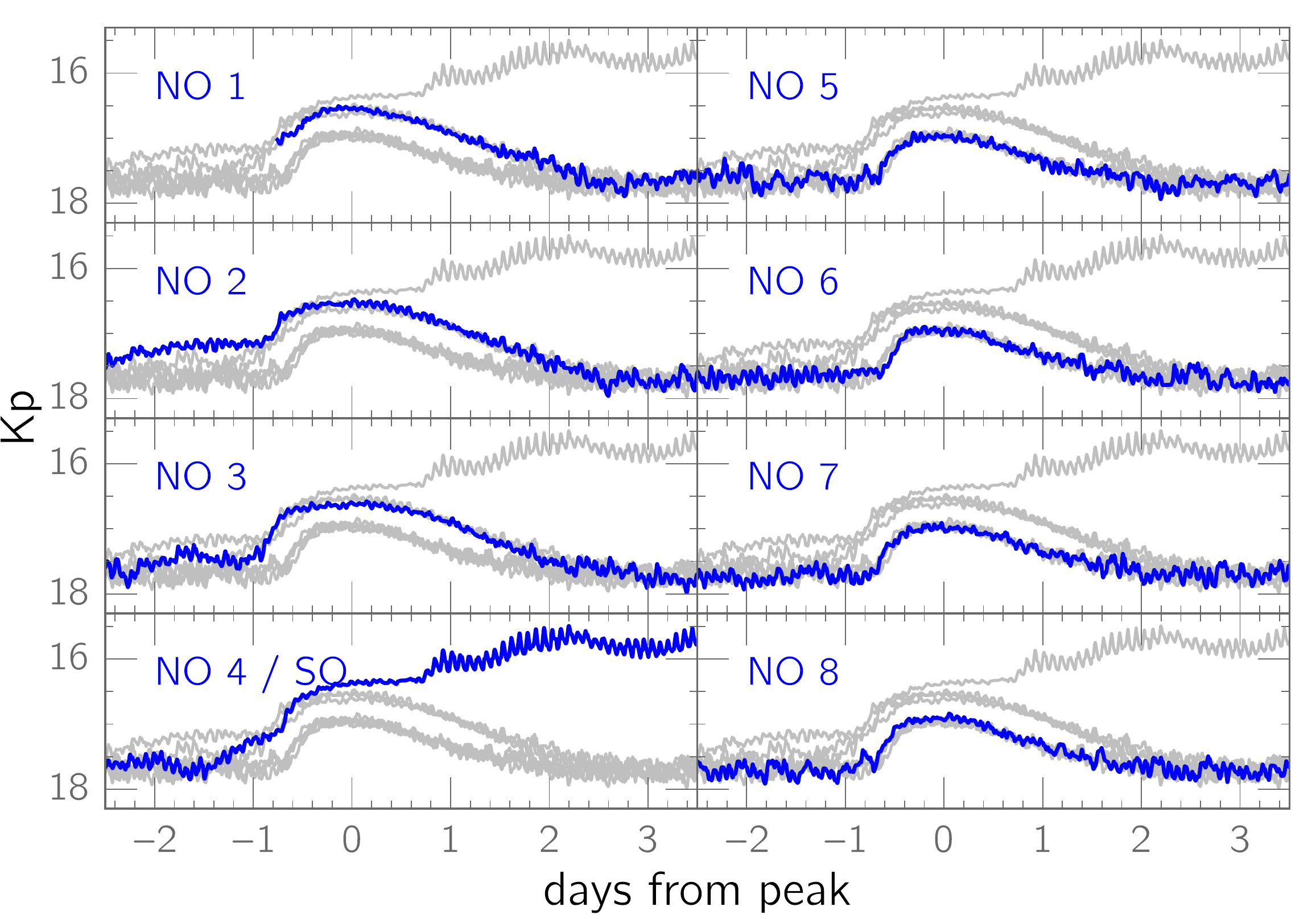}
\caption{An overlay of smoothed light curves of the normal outbursts. The peak of each outburst is centered on t = 0 days, and each panel highlights a different normal outburst. Before the superoutburst (left column), there was scatter in the peak brightnesses of the different outbursts, and the quiescent level prior to outburst was variable as well. After the superoutburst (right column), the normal outbursts peaked at similar magnitudes, had comparable rise and decline times, and showed similar quiescent magnitudes in the lead-up to the outburst. The precursor of the superoutburst was brighter than the other normal outbursts, and it was still brightening at the same time that the other outbursts were declining.
\label{normal_outbursts_fig.pdf}}
\end{figure*}

\begin{figure}
\includegraphics[width=\columnwidth]{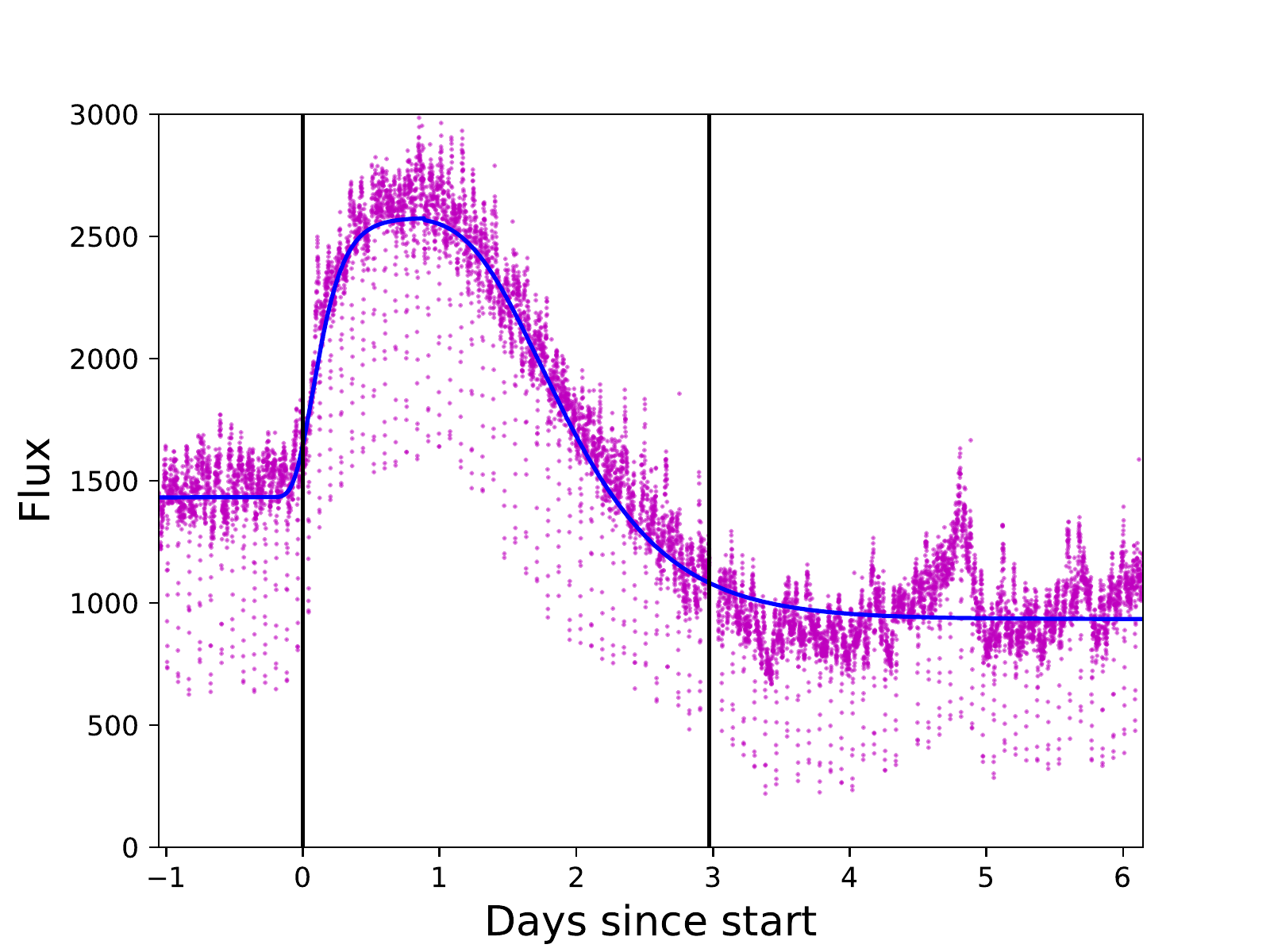}
\caption{The result of fitting equation \ref{step_func} (blue) to the second outburst in the \textit{K2} light curve (magenta). The black lines denote the times at which the model flux reach 1.15 times the value of the constant level. The pre-outburst flux level before the outburst was much higher than the quiescent level afterwards. \label{burst3}}
\end{figure}

There are a total of eight normal outbursts in the \textit{K2} data, including the precursor normal outburst. Table ~\ref{tab:burst parameters} lists the details of these outbursts, and Fig.~\ref{normal_outbursts_fig.pdf} overlays each of the normal outbursts so that their amplitudes and shapes may be compared. They were modeled using a pair of Gompertz functions such that the rise and fall could be described by separate Gompertz functions, joined in the middle. The function describing the flux took the form 
\[
    F(t)= 
\begin{cases}
    a_1\:e^{-e^{-k_{1}(t-t_1)}}+c_1,& \text{for } t\leq t_0\\
    a_2\:e^{-e^{-k_{2}(t-t_2)}}+(c_1+a_1),& \text{for } t\geq t_0\\
\end{cases}\label{step_func}
\]
Here, $a_1$ and $a_2$ are the amplitude of the rise and fall of the outburst, $k_1$ and $k_2$ respectively control how fast the rise and fall of the outburst were, $c_1$ is the quiescent level before the outburst, $t_1$ is the mid time of the rise, $t_2$ is the mid time of the fall, and $t_0$ is the time at which the function transitions from rising to falling. Allowing different $a_1$ and $a_2$ values for a single outburst enables the quiescent brightness to change after the outburst. Table~\ref{tab:burst parameters} lists the model parameters obtained by fitting this function to each normal outburst, and Fig.~\ref{burst3} shows a representative fit to one of the outbursts.

\begin{table*}
	\centering
	\caption{The parameters from fitting equation \ref{step_func} to each of the visible outbursts. The superoutburst parameters (SU in the table) represent the parameters for the precursor of the superoutburst. Due to the strong orbital variations in the light curve caused by the bright spot and eclipse, we have taken the error in the timing measurements to be the length of a single orbit.}
	\begin{tabular}{c c c c c c c}
		\hline
		Outburst 	& Start Date 					& Duration      & Time between                                      & Amplitude\tablenotemark{b}    & Rise time                 & Decay time \\
		            & (BJD)                         & (days)        & outbursts (days)\tablenotemark{a}                 &                               & (days)                     & (days) \\
		\hline\hline
	    1           & 2457061.91(8)                 & 2.51(8)       & -                                                 & 2.6(1)\tablenotemark{c}       & $>$0.56\tablenotemark{d}  & 1.96(8)     \\
	    2           & 2457071.02(8)                 & 2.90(8)       & 9.12(8)                                           & 2.8(1)\tablenotemark{c}       & 0.92(8)                   & 1.98(8)     \\
	    3           & 2457079.38(8)                 & 3.20(8)       & 8.36(8)                                           & 2.3(1)                        & 1.15(8)                   & 2.05(8)     \\
	    4 (SU)          & 2457089.94(8)                 & 1.86(8)       & 10.56(8)                                          & 3.2(1)                        & 1.73(8)                   & -           \\
	    5           & 2457105.48(8)                 & 2.12(8)       & 15.54(8)                                          & 2.0(1)                        & 0.74(8)                   & 1.37(8)     \\
	    6           & 2457112.41(8)                 & 2.24(8)       & 6.93(8)                                           & 1.9(1)                        & 0.55(8)                   & 1.69(8)     \\
	    7           & 2457118.78(8)                 & 2.24(8)       & 6.37(8)                                           & 2.0(1)                        & 0.50(8)                   & 1.75(8)     \\
	    8           & 2457126.21(8)                 & 2.31(8)       & 7.42(8)                                            & 2.2(1)                        & 0.86(8)                   & 1.44(8)     \\
		\hline
	\end{tabular}
	\begin{flushleft}
	\textit{
	\tablenotetext{a}{ The time between outbursts is defined as the time between the start of the given outburst and the start of the previous outburst. Because of this, outburst 1 does not have a value.}
	\tablenotetext{b}{Amplitudes are expressed as flux ratios between the maximum flux during outburst and the pre-outburst quiescent flux.}
	\tablenotetext{c}{The amplitudes of these outbursts were calculated using the quiescent flux after the outburst had ended. This was due to the lack of a quiescent level before outburst 1 (since the outburst was rising at the start of the exposure) and due to an abnormally high pre-outburst level for burst 2 (see the top panel of Fig.~\ref{burst3}).}
	\tablenotetext{d}{We can only provide a lower bound for the rise time of the burst since outburst 1 started before the beginning of the \textit{K2} observations.}
	}
	\end{flushleft}
	\label{tab:burst parameters}
\end{table*}

The parameters in Table~\ref{tab:burst parameters} show that before the superoutburst, the amplitude of the normal outbursts was $\sim$0.4~mag larger, and the fade to quiescence took about a half-day longer. Moreover, the fast rise times and comparatively slow decay rates indicate that the outbursts originated in the outer disk and moved inward. The behavior of the eclipse width, which reached a maximum almost immediately after the onset of the normal outbursts, supports this interpretation. An outburst starting in the outer disk will cause an immediate jump in the eclipse width because the donor star cannot eclipse the entire outbursting region of the disk. By contrast, in the early stages of an inside-out outburst, the outburst luminosity would be confined to the inner disk, which would be eclipsed more quickly and more completely than the outer disk, resulting in deeper, narrower eclipses \citep{webb}.

During some of the normal outbursts (particularly the sixth and eighth normal outbursts), the time of eclipse minimum occurred up to $\sim$15~sec earlier than in quiescence (Fig.~\ref{full_LC}). This phenomenon probably resulted from the diminished relative contribution of the hotspot to the overall light curve during an outburst. Because the hotspot is eclipsed after the center of the disk, the eclipses will occur later if the hotspot is dominant, as is the case during quiescence. During outburst, the disk becomes more luminous than during the quiescent state, so the disk's centroid of emission is eclipsed at an earlier orbital phase \citep{ram17}.

The properties of eclipses during the normal outbursts change after the superoutburst, becoming narrower in FWHM by $\sim$30 sec and deeper by several tenths of a magnitude. These observations imply that the maximum radius of the disk during outburst was smaller after the superoutburst, a key prediction of the TTI model. Although we attempted to use the ELC code to quantitatively measure the change in disk radius, there were too many free parameters to achieve a reliable measurement.

\begin{figure*}
\includegraphics[width=\textwidth]{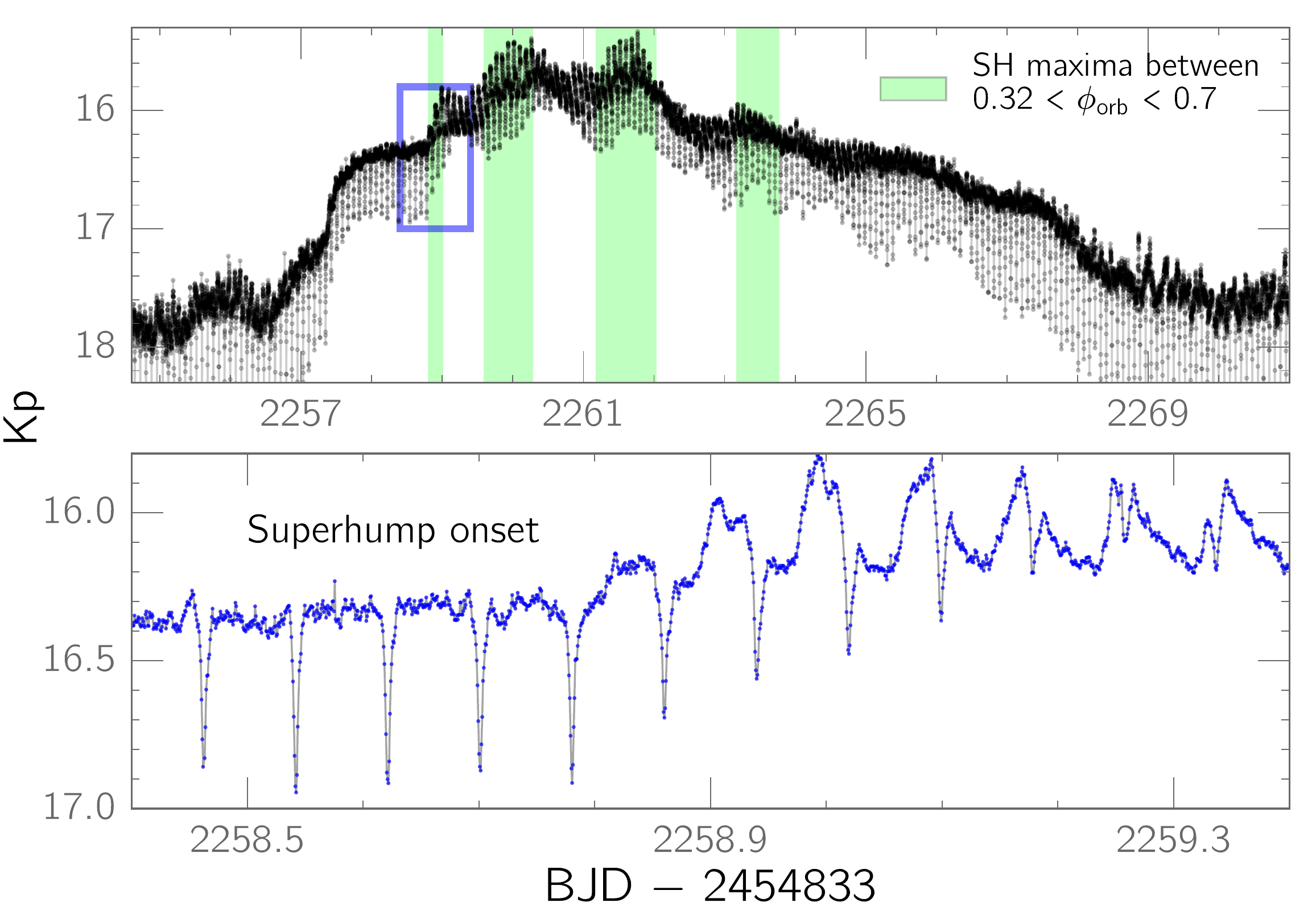}
\caption{The light curve of the superoutburst. The segment enclosed by the rectangle indicates where superhumps appeared, and the bottom panel shows an enlarged view of this segment of the light curve. The highlighted regions indicate parts of the superoutburst during which the superhump amplitude was largest, and as discussed in the text, there is an orbital-phase dependence to this phenomenon. \label{superhump_fig}}

\end{figure*}

\subsection{Precursor Outburst}

At approximately BJD 2457090.37, \mls\ initiated a precursor normal outburst that subsequently triggered a superoutburst. The rise to precursor maximum took approximately 0.7 days, and the superhumps appeared an additional $\sim$0.7 days thereafter, a total of $\sim$18 orbital cycles from the start of the precursor.

Compared with the three previous normal outbursts, the precursor was $\sim$0.2 mag brighter at maximum light, and whereas the other outbursts declined after reaching peak brightness, the precursor plateaued after its maximum. Immediately after the precursor's peak, the eclipse FWHM began to decline and the eclipse depth increased, exactly as was observed in the other normal outbursts (Fig.~\ref{full_LC}). Both effects are consistent with the inward propagation of a cooling front, but the lack of a concomitant fade in the overall brightness implies the presence of an additional mechanism to offset the cooling front's reduction of the disk luminosity.

Moreover, in the three orbits prior to superhump onset, the hotspot was no longer discernible. During these orbits, there were several prominent dips of unknown origin, and the overall brightness gradually increased by $\sim$0.1 mag.

\subsection{Superoutburst light curve} \label{SO_sec}

The first superhump appeared at BJD$\sim$2457091.82, at which time the light curve immediately began rising to superoutburst maximum. Once the superhumps appeared, their development was very rapid, and by the third superhump cycle, their amplitude had already reached its maximum value of 0.4~mag (Figure~\ref{superhump_fig}, lower panel). The superhump amplitude might have been even larger, but the superhumps and eclipses beat against each other, distorting the profiles of many superhump maxima. The rise to superoutburst maximum lasted until BJD$\sim$2457093.3, although the presence of eclipses and superhumps makes it difficult to reliably discern the exact time at which the superoutburst peaked.

\begin{figure*}
\epsscale{1.18}
\plotone{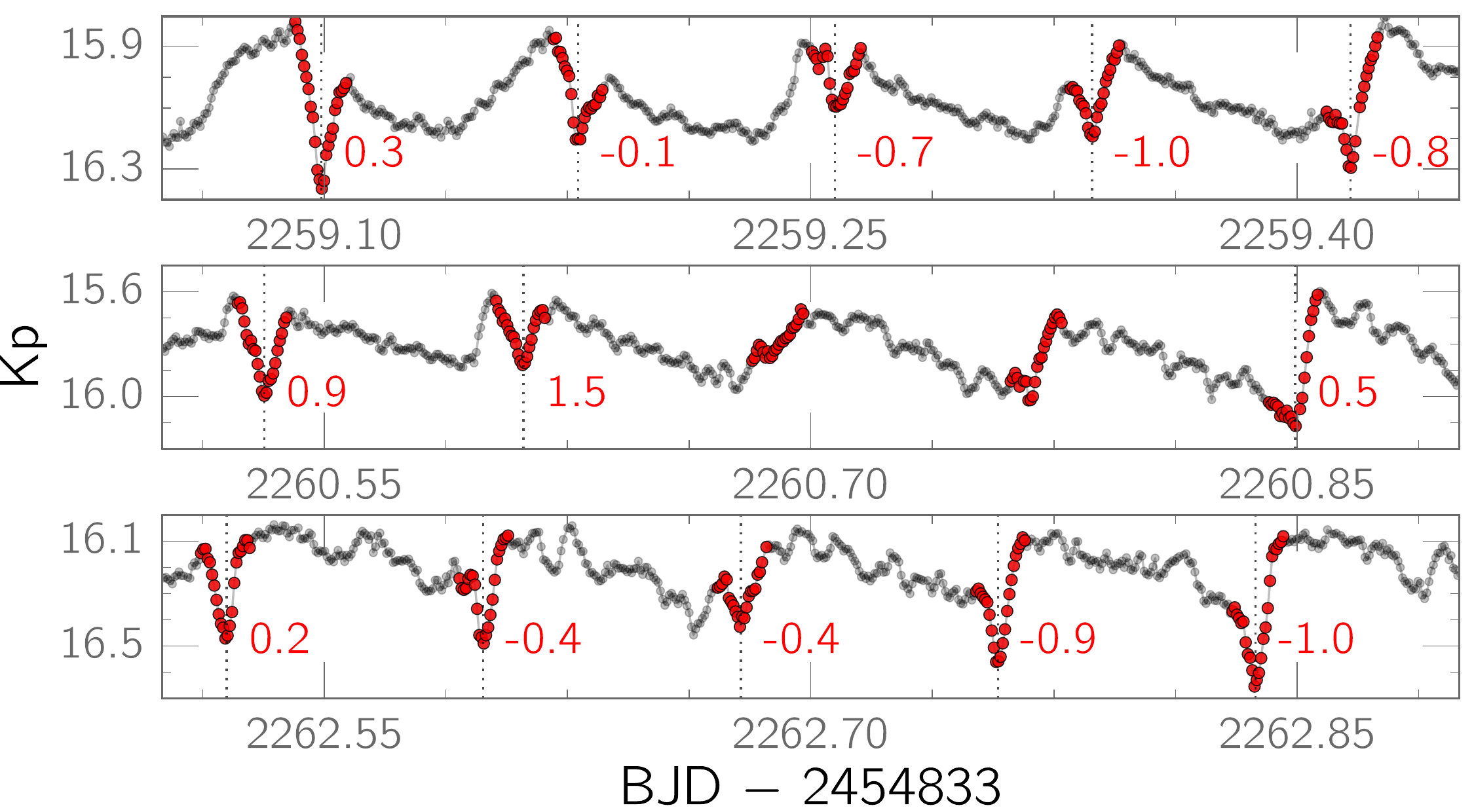}
\caption{Light curves showing three different times at which the eclipses coincided with the superhump maxima. The red, enlarged points denote data obtained between $ 0.9 < \phi_{orb} < 1.1$, and the red number next to each eclipse is the O$-$C timing residual for mid-eclipse, given in minutes. The dotted vertical lines indicate the measured time of minimum light. No value is given for two eclipses in the middle panel because those eclipses are essentially indistiguishable from the superhump profile. \label{superhumps_and_eclipses}}
\end{figure*}


During the superoutburst, the eclipse O$-$C timings (Fig.~\ref{superoutburst_o-c}) showed oscillations when the eclipses lined up with the superhump light source. Fig.~\ref{superhumps_and_eclipses} shows three segments of the superoutburst light curve during which the eclipses occurred at superhump maximum. For example, when the eclipses and superhump maxima coincided during the initial rise to superoutburst maximum, mid-eclipse occurred approximately one minute earlier than predicted by the orbital ephemeris (top panel in Fig.~\ref{superhumps_and_eclipses}). This implies that the superhump light source was located in the trailing half of the disk. Later, near BJD$\sim$2457093.8, two consecutive eclipses were so shallow that they could not be readily distinguished from the superhump profile, implying a grazing eclipse of the superhump light source. 

Another noteworthy aspect of the superoutburst is that it contained several shallow fading events during which the overall brightness decreased by several tenths of a magnitude. These dips can be seen in the gaps between the four highlighted regions in the top panel of Fig.~\ref{superhump_fig}. The fades lasted for about one day, and each was centered on the time that the eclipses coincided with the superhump maxima.

\begin{figure*}
\epsscale{1.22}
\plotone{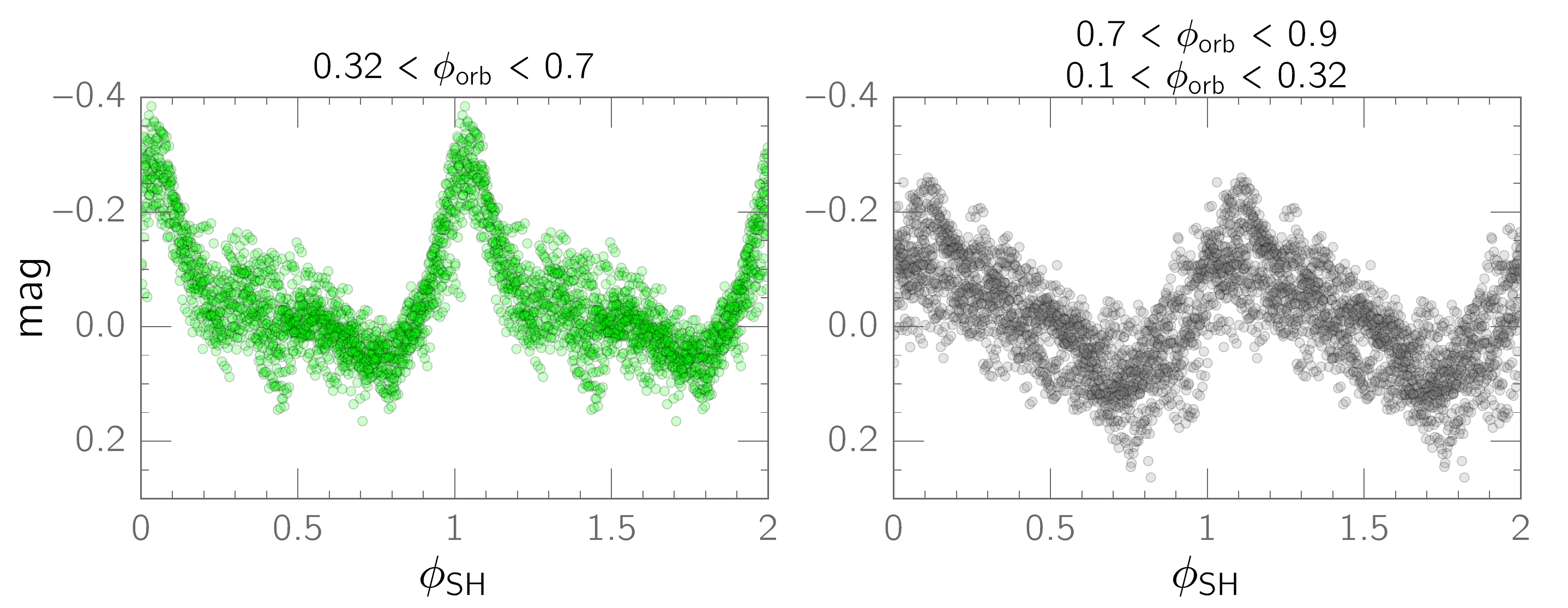}
\caption{Reconstructed profiles of Stage B superhumps between BJD 2457092.8-2457096.0, when the superhump period was stable at 2.003~h. We built the profiles using the detrended light curve described in the text, and then filtered the data by orbital phase. Superhumps observed between orbital phases 0.32-0.7 had a sharper profile and larger amplitude than those observed at other orbital phases. Observations between orbital phases 0.9-1.1 were excluded from these plots because of unavoidable eclipse contamination. 
\label{superhump_orbital}}

\end{figure*}

The fading events showed a strong dependence on the orbital phase at which the superhump maximum occurred. After subtracting the smoothed light curve to isolate the pulsed flux, we constructed two phase plots of the Stage B superhump profile, when the superhump period was constant at 2.003~h. The first phase plot used only data obtained between $0.32 < \phi_{orb} < 0.7$, which, from visual inspection of the light curve, we estimated to be the orbital phases during which the superhump maximum was the brightest and most clearly defined. The second phase plot of the superhump used data from all remaining orbital phases (except $ 0.9 < \phi_{orb} < 1.1$, as these phases are inevitably contaminated by eclipses). The resulting phase plots are shown in Fig~\ref{superhump_orbital}. The reconstructed superhump profile for $0.32 < \phi_{orb} < 0.7$ is sharper and has a higher amplitude than the noisy superhump profile for the remainder of the data.

There are at least two possible explanations for this phenomenon. \citet{om03} calculated that in high-inclination systems, the strength of the superhump signal will depend on the orbital phase at which the superhump occurs. According to their work, about half of the superhump light originates in the vertically-extended rim of the disk, the visibility of which will vary across the orbital cycle when the disk is seen edge-on. In their Figures 2 and 3, they predict that superhump maxima will be strongest for about half of an orbit centered on an orbital phase of $\sim$ 0.75, mimicking the orbital hump from the stream-disk hot spot. This differs from the orbital phases (0.32-0.70) at which we observed the highest-amplitude superhumps in \mls.

Another possible cause of the fades is obscuration of the inner disk by vertically extended disk structure at certain orbital phases. In this scenario, the superhump light source would have been partially obscured except when it occurred between $0.32 < \phi_{orb} < 0.7$. \citet{billington96} reported evidence of such structure in the eclipsing system OY~Car, in which the optical superhump maxima corresponded with deep ultraviolet dips. Furthermore, \citet{murray98} predicted that the temperature gradient of a superhumping disk could be capable of producing azimuthal variations in the height of the disk rim.

\subsection{Period of Stage A superhumps}
\label{StageA_sec}

\begin{figure}
\includegraphics[width=\columnwidth]{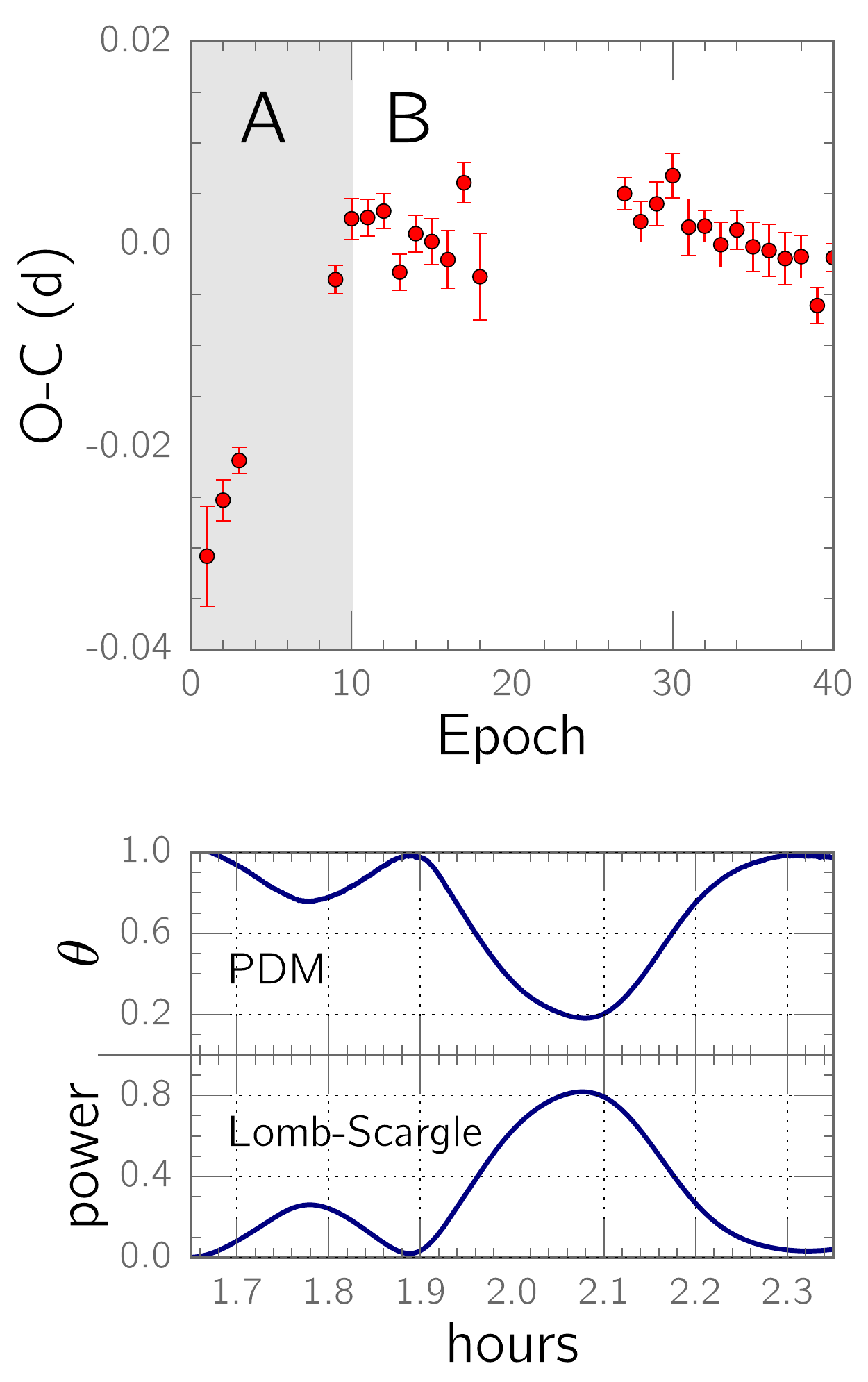}
\caption{{\bf Top:} An O$-$C diagram of the Stage A-B superhump transition, showing a Stage A period of $2.089\pm0.007$~h. The times of maxima were measured by fitting superhumps with polynomials, and maxima between $0.9 < \phi_{orb} < 1.1$ are not plotted because of eclipse contamination. {\bf Bottom:} PDM and Lomb-Scargle periodograms of Stage A, yielding periods of 2.076$\pm$0.005~h and 2.078$\pm$0.005~h, respectively. Before they were computed, observations between $0.9 < \phi_{orb} < 1.1$ were excluded, and the light curve was detrended. Candidate periods are minima with PDM and maxima with Lomb-Scargle. \label{stageA_fig}}
\end{figure}

To measure the period of Stage A superhumps, we performed an O$-$C analysis (Fig.~\ref{stageA_fig}) by fitting the superhump maxima with polynomials and extracting the time of maximum light. All observations between $0.9 < \phi_{orb} < 1.1$ were excluded to reduce the effect of the orbital modulation. A robust Theil-Sen linear fit to the timings of the Stage A maxima shows the superhump period to have been constant at $2.089\pm0.007$~h before an abrupt jump to a 2.003~h period, consistent with a transition from Stage A to B superhumps. Five of the ten stage A superhumps were adulterated by eclipses, so their maxima were not included in the analysis. To quantify the timing uncertainties on individual maxima, we performed 100 simulations in which we repeated the fitting procedure after randomly selecting half of the data points that had been used to fit each individual superhump. We then used a Monte Carlo simulation to estimate the uncertainty of the Stage A period.

We also calculated Lomb-Scargle and phase-dispersion-minimization \citep[PDM; ][]{PDM} periodograms for Stage A after subtracting the smoothed light curve (Fig.~\ref{full_LC}) to detrend the data. We excluded observations between $0.9 < \phi_{orb} < 1.1$ after determining that their inclusion systematically inflated the measured period of Stage A. Using two harmonic terms, the Lomb-Scargle analysis yielded a period of 2.076$\pm$0.005~h, while the period in the PDM periodogram was 2.078$\pm$0.005~h. These periodograms are shown in Fig.~\ref{stageA_fig}. To estimate the 1$\sigma$ uncertainties, we used a method similar to one from \citet{kato12}. We performed the Lomb-Scargle and PDM analyses on 100 random subsets, each of which contained half of the observations during Stage A, and we derive the uncertainties from the resulting distributions.

We caution, however, that the overlap between eclipses and superhump maxima can bias the measured period of Stage A, corrupting the computed mass ratio \citep[as occurred during the 2010 superoutburst of HT Cas; ][]{ko13}. Of the three methods that we used, the O$-$C analysis of the polynomial fits to the superhump maxima is probably the most vulnerable to this problem, so we excluded Stage A maxima that fell between $0.9 < \phi_{orb} < 1.1$. Since the Lomb-Scargle and PDM periodograms use the full superhump profile---and not just the maximum---they might be less susceptible to distortions in the profiles of individual superhump maxima. 

In Sec.~\ref{mass_ratio_sec}, we use these three estimates of the Stage A period to calculate the corresponding binary mass ratio.

\subsection{Phase shift of the hotspot}
\label{hotspot_sec}

\begin{figure}
\epsscale{1.2}
\plotone{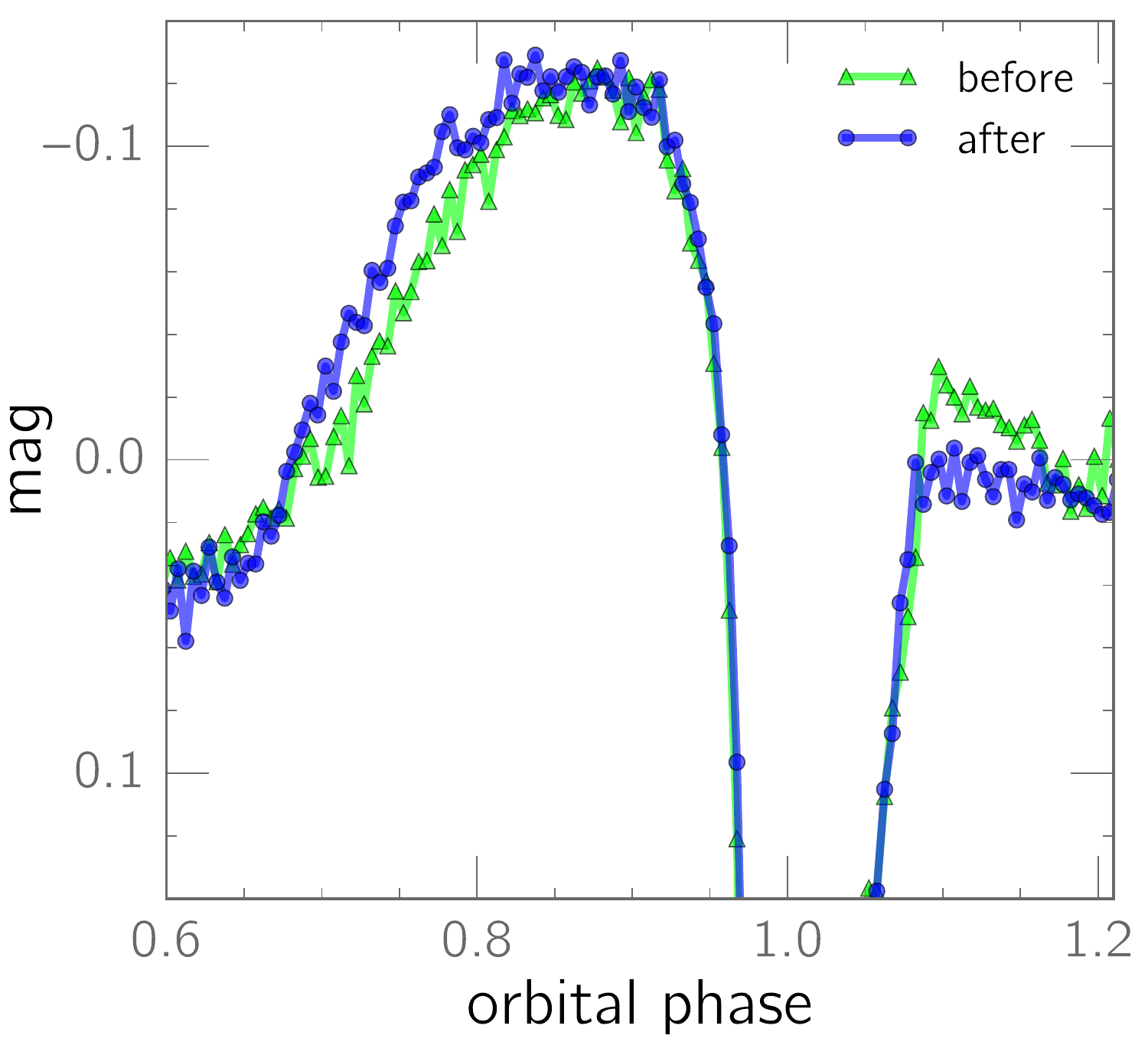}
\caption{A phased light curve of the hotspot, before and after the superoutburst. Before the data were phased, they were detrended by subtracting a smoothed light curve. The bin width for each point is 0.005 phase units. The orbital hump shifted towards earlier phases after the superoutburst, as did the hotspot's egress feature. \label{hotspot_fig}}
\end{figure}

Using the detrended light curve, we created phase plots of the quiescent orbital modulation before and after the superoutburst. A comparison of these plots, shown in Fig.~\ref{hotspot_fig}, reveals that the hotspot shifted towards earlier orbital phases after the superoutburst. The significance of this effect is discussed in Sec.~\ref{disk_size}.

\section{Spectroscopy} \label{spectra_sec}

\begin{figure*}
\epsscale{1.2}
\plotone{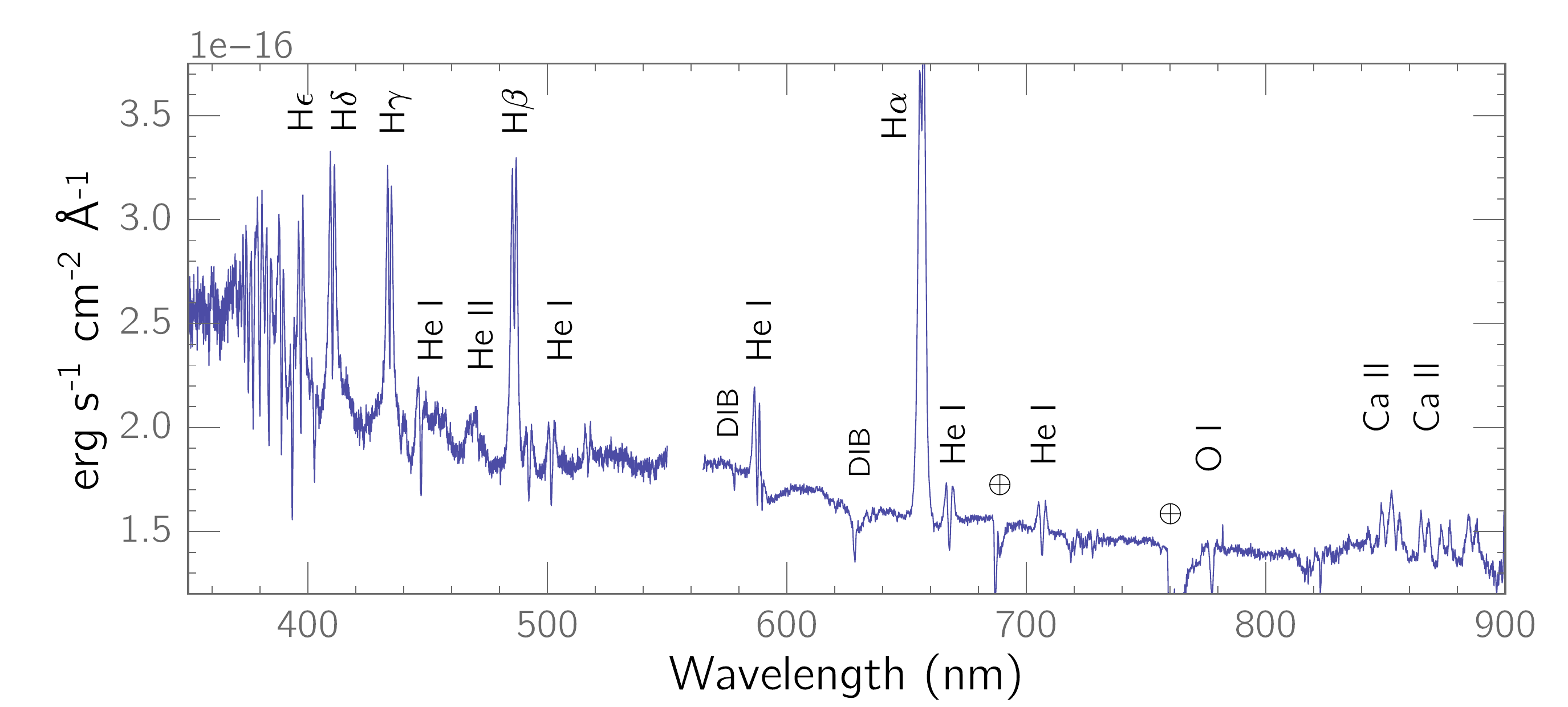}
\caption{Average spectrum of J0359 in 2016, with major lines identified. ``DIB'' refers to diffuse interstellar bands. The Balmer lines blueward of H$\delta$ and the He I lines all show central absorption components. O I $\lambda$ 777~nm is seen in absorption. The gap near 560~nm is due to a dichroic that splits the light into the two MODS spectrographs. No velocity correction has been applied. \label{spectrum}}
\end{figure*}

The 2015-2016 LBT and 2014-2015 APO spectra were largely comparable, except for minor variations in the continuum slope. All spectra showed double-peaked Balmer and He I emission lines from the disk, with the centers of the He I lines dipping below the continuum into absorption (Fig.~\ref{spectrum}). There was also weak He~II $\lambda4686$ \AA\ emission.

O I $\lambda7774$ \AA\ was seen in absorption with a FWHM of $\sim500$ km s$^{-1}$, exhibiting radial-velocity variations on the orbital period. Its phasing suggests an origin within the disk, and unlike the disk emission lines, it disappeared entirely during the eclipse.

\begin{figure*}
\epsscale{1.0}
\plottwo{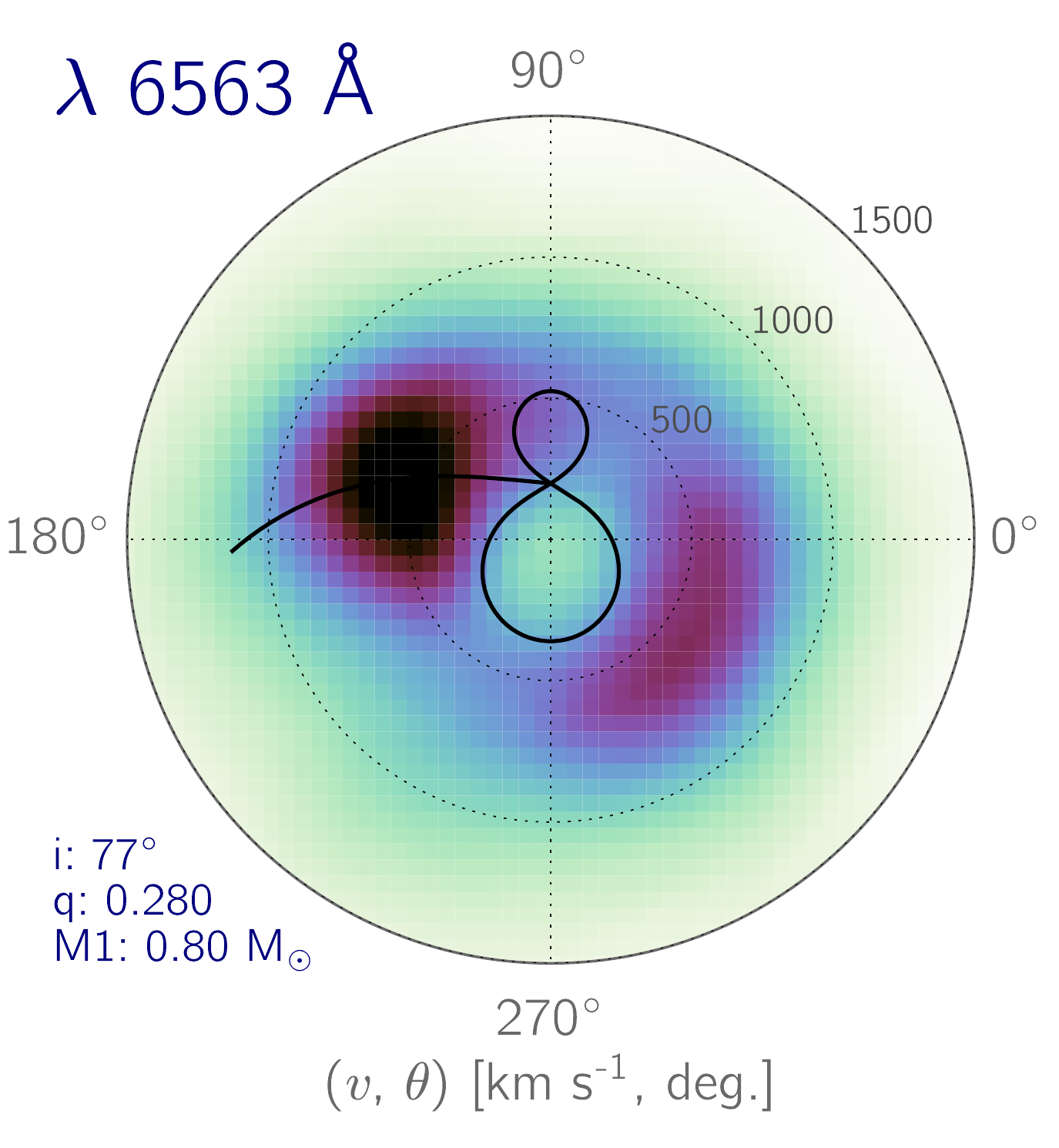}{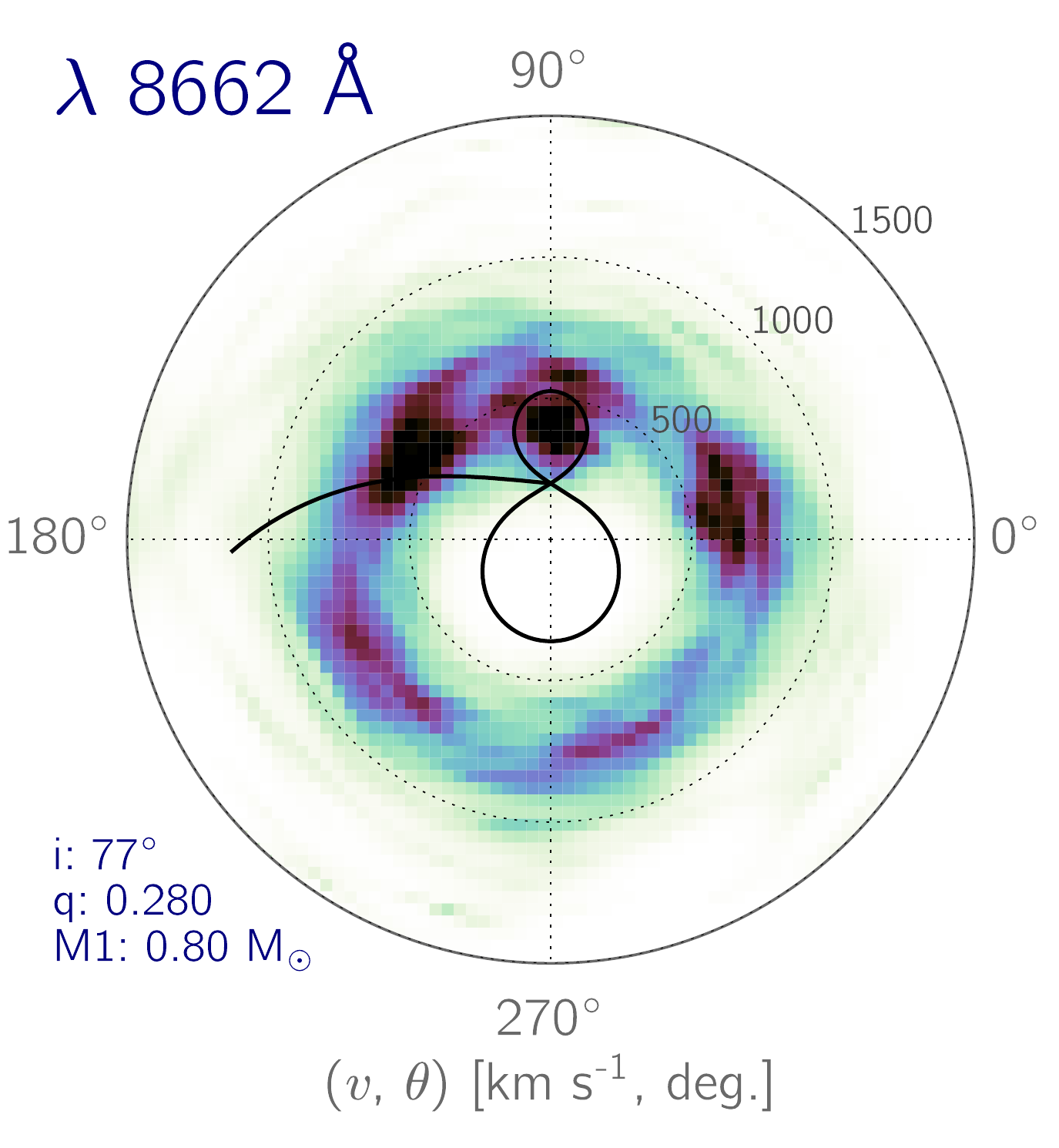}
\caption{Doppler tomograms of H$\alpha$ (left) and Ca II (right). The WD mass has been assumed. Emission from the donor was present only in the Ca II lines. The Roche lobes corresponding to $q=0.28$ have been plotted, as has the ballistic trajectory of the accretion stream.
\label{tomograms}}
\end{figure*}

We constructed Doppler tomograms from the H$\alpha$ and Ca II $\lambda 8662$ \AA\ lines in the 2016 LBT data using code by \citet{kotze}, and we show them in Fig.~\ref{tomograms}. Both tomograms are dominated by the stream-disk hotspot, and the Ca II tomogram also shows weak emission from the donor.

\subsection{Radial velocity of the donor star}

Emission from the donor star was detectable in the Ca II $\lambda\ 8542$ \AA\ and $\lambda\ 8662$ \AA\ lines, but the former overlapped with emission from another Ca II line. Thus, we used the $\lambda$ 8662 \AA\ line to characterize the orbital motion of the donor. A trailed spectrum of this line shows double-peaked emission from the disk with a partial S-wave superimposed. The phasing of the S-wave, with maximum redshift at $\phi_{orb} = 0.25$ and maximum blueshift half a cycle later, clearly indicates that it originated on the donor. The S-wave was apparent (albeit weakly) for only half of the orbit before it became overwhelmed by disk emission. Although we searched for additional spectral lines from the secondary, such as the near-infrared K I and Na I absorption doublets, we detected only the two Ca II lines.

We used two methods to constrain the semiamplitude ($K_{2}$) and systemic velocity ($\gamma$) of the donor based on the Ca II emission. First, we visually fitted a sinusoid with a period of 1 to the trailed spectrum, adjusting $K_{2}$ and $\gamma$ to achieve a satisfactory fit. Second, we applied a two-dimensional cross-correlation of the trailed spectrum, with the template being half of a sine wave with a fixed period of 1. We iterated across a range of plausible values for $K_{2}$, rebuilding the template for each such value. Both methods agree that $K_{2} \approx$ 320 km s$^{-1}$ and $\gamma \approx 20$ km s$^{-1}$. We do not have formal 1-$\sigma$ uncertainties for these values, but we visually estimate uncertainties of $\pm$20 km s$^{-1}$ for both $K_{2}$ and $\gamma$.

A limitation of this method is that it assumes that the observed emission tracks the donor's center of mass. In reality, there is no such guarantee, as different regions on the donor will have different orbital speeds, depending on their location within the donor's Roche lobe. The absence of the donor's emission lines during the eclipse shows that the lines were produced on the donor's inner hemisphere, most likely as the result of irradiation of regions that were unshielded by the accretion disk. Consequently, the observed donor-star emission probably originated somewhere between the L1 point and the secondary's center of mass, in which case our measurement of $K_{2}$ underestimates the true value for the donor.

\subsection{Radial velocity of the disk}

We estimated the disk's velocity variations in the LBT data by using the ``double Gaussian'' method developed by \citet{shafter83}. The method consists of convolving an emission line with two Gaussian functions separated in wavelength. The wavelength at which the Gaussian functions contain equal flux is an estimate of the velocity centroid of the emission from the inner disk, approximating the motion of the WD. We used a Gaussian sigma of 7~\AA, which is typical for this analysis. We analyzed the bright, uncontaminated H$_\alpha$ and H$_\beta$ emission lines, and their velocities were fitted to the function

\begin{equation}
V(t) = -K_1\;\sin\left(\frac{2\pi(t-t_0)}{P} +\phi_0\right) + \gamma,
\end{equation}

with the free parameters of WD velocity amplitude ($K_1$), velocity offset ($\gamma$), and phase ($\phi$). We fixed the orbital period ($P$) to that derived from the \textit{K2} photometry. This process was repeated for a range of Gaussian separations until the parameters that provided the minimum velocity scatter were found. The velocity estimates made around the eclipse were dominated by the Rossiter-McLaughlin effect, so their errorbars have been inflated to avoid strongly influencing the velocity fits.

\begin{figure}
\includegraphics[width=0.5\textwidth]{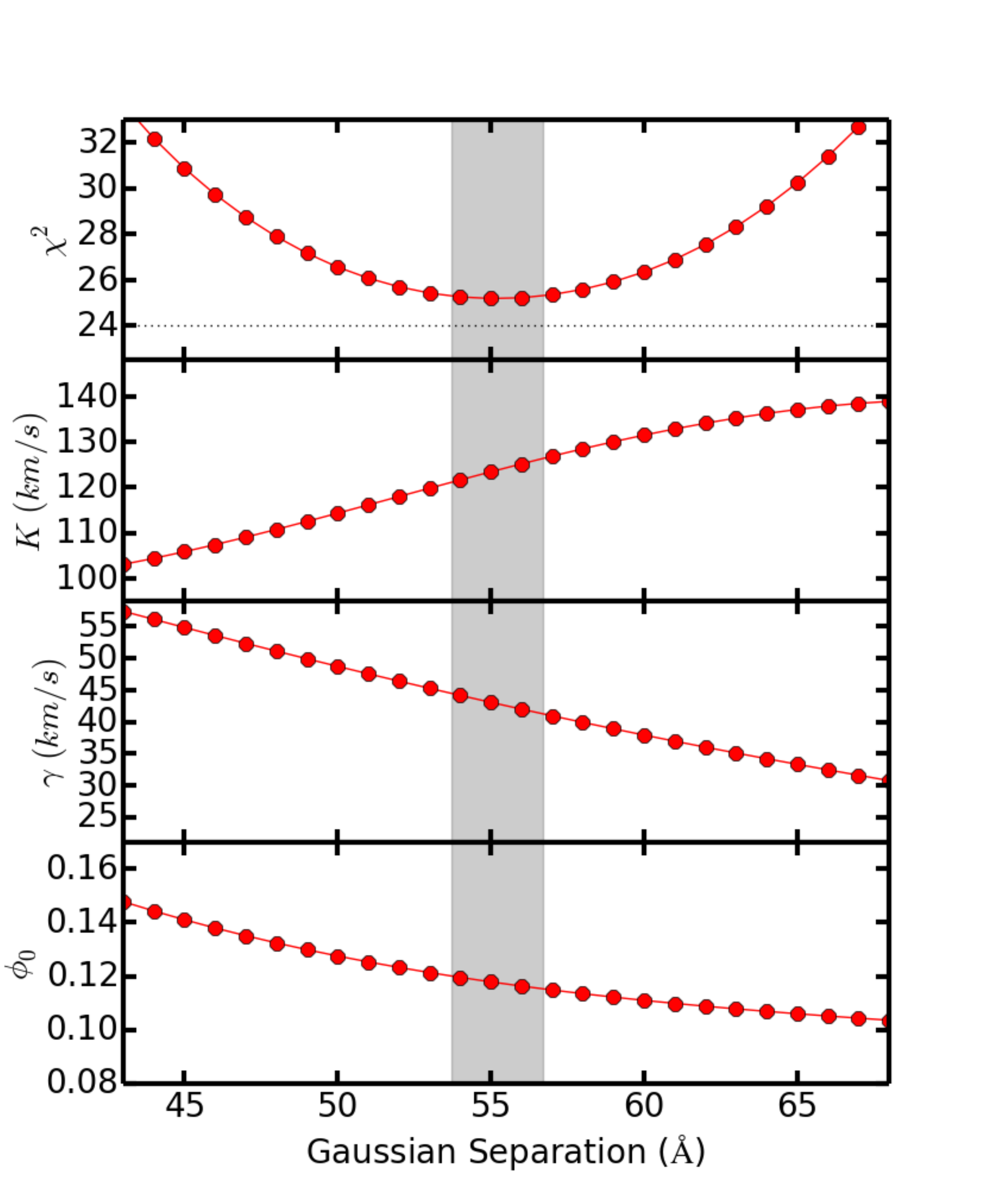}
\caption{The double-Gaussian method applied to the H$_\alpha$ emission feature, showing that the lowest scatter occurs for a Gaussian separation of 55~\AA. The velocity amplitude and center of mass velocity vary slowly with the separation
parameter. The time of zero radial velocity shows a significant offset from
the time of the photometric eclipse with $\phi_0 = 0.12$.  \label{shafter}}
\end{figure}

\begin{figure}
\includegraphics[width=0.5\textwidth]{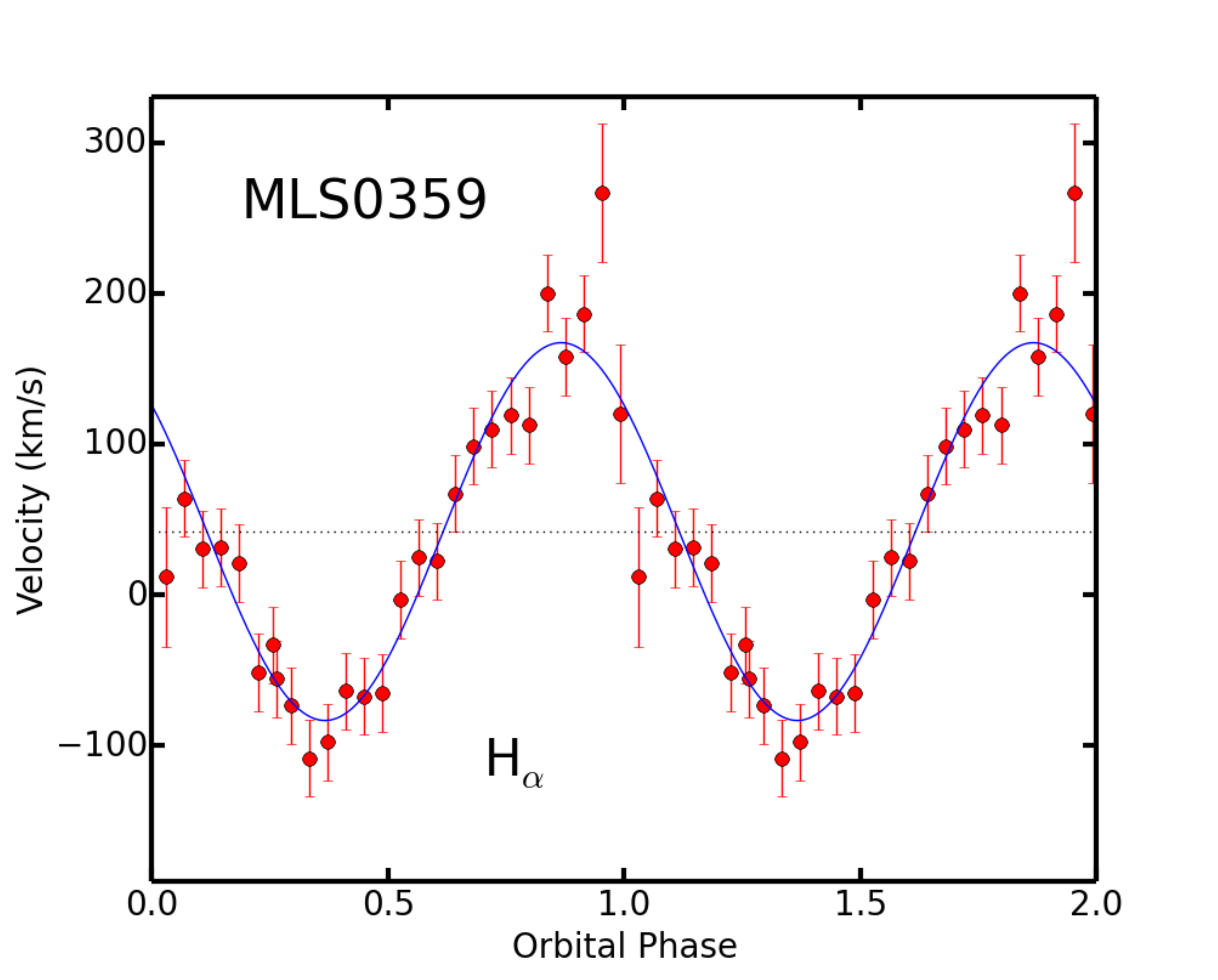}
\caption{Radial velocity curve of the inner disk, approximating the motion of the WD. The orbital phase is based on the time of the photometric eclipse. The three spectroscopic measurements during eclipse have been discounted from the radial velocity fit by inflating their uncertainties.  \label{vel_amplitude}}
\end{figure}

Application of the double-Gaussian method to the H$_\alpha$ emission (Fig.~\ref{shafter}) results in $K_{1}=123\pm 4$ km~s$^{-1}$ and $\gamma = 43\pm 3$ km~s$^{-1}$. The offset in phase between the best-fit sinusoid and the time of eclipse is $\phi_0 = 0.12\pm 0.01$ (Fig.~\ref{vel_amplitude}). The same analysis applied to the H$_\beta$ emission yields $K_{1}=129\pm 5$ km~s$^{-1}$ and $\phi_0 = 0.12\pm 0.02$, consistent with the H$_\alpha$ results. The time-resolved APO spectra yield similar values, showing the disk to be relatively stable over timescales of years.

\section{Discussion}

\subsection{Mass ratio}
\label{mass_ratio_sec}

There is a fundamental relationship between a CV's mass ratio ($q = \frac{M_2}{M_1}$) and its ability to develop superhumps. The TTI model predicts that superhumps form when the outer disk achieves a 3:1 resonance with the donor star, but if the mass ratio is too high, tidal forces from the secondary truncate the disk before it can reach this size. In simulations, superhumps do not appear if $q \gtrsim 0.25$ \citep{smith07}.

As discussed in Sec.~\ref{introduction}, \citet{ko13} proposed that the period of Stage A superhumps is equivalent to the dynamical precession rate at the 3:1 resonance, meaning that the mass ratio  can be very accurately determined by measuring the fractional excess ($\epsilon^{*} = 1- \frac{P_{orb}}{P_{sh}} $) of Stage A superhumps. In Sec.~\ref{StageA_sec}, we used three different techniques to measure the period of Stage A, obtaining three estimates of the period. For each, we measured $\epsilon^{*}$ using the orbital period of 1.909~h and applied Eq. 1 in \citet{ko13} after correcting for a misprint \citep[][ footnote 5]{kato16}.

The O$-$C, PDM, and Lomb Scargle periods were 2.089$\pm$0.007~h, 2.078$\pm$0.005~h, and 2.076$\pm$0.005~h, respectively, and the corresponding period excesses result in mass ratios of 0.298$\pm$0.016, 0.275$\pm$0.011, and 0.270$\pm$0.011. These uncertainties assume that the uncertainty on each period is the standard deviation of a Gaussian distribution whose mean is the measured period. The average of these mass ratios is $q = 0.281\pm0.015$, where the uncertainty is the RMS of the three estimates of the mass ratio. This value is marginally inconsistent with the theoretically predicted threshold of $q = 0.25$ for superhump development, though we reiterate that the contamination of Stage A superhump maxima by eclipses can potentially lead to an inaccurate measurement of the Stage A period \citep{ko13}. At the very least, \mls\ is very close to $q = 0.25$, but it will take an independent measurement of the mass ratio to conclusively establish whether $q > 0.25$.

Because we have estimated $K_{1}$ and $K_{2}$ from the spectra, we have a second means of nominally estimating the mass ratio. For $K_{1} = 123$ km s$^{-1}$ and $K_{2} = 320$ km s$^{-1}$, the mass ratio would be $q$ = 0.38. However, it is likely that there are systematic errors impacting both values. Given the caveats described earlier, our value for $K_{2}$ is probably a lower limit for the true orbital motion of the secondary. Moreover, as \citet{lg99} showed, estimates of $K_{1}$ from the line wings can deviate significantly from the true orbital motion of the WD. While there is insufficient data to obtain a reliable dynamical estimate of the mass ratio, the large changes that are needed to satisfy $q = \frac{K_{1}}{K_{2}} \simeq 0.25$ furnish modest support for the high mass ratio implied by the superhump method.

Although \citet{mww00} identified a scenario in which superhumps could develop in systems with mass ratios as high as $q = 0.33$, their proposal requires the mass-transfer rate to abruptly plummet (e.g., as in VY~Scl stars). However, the closely spaced outbursts in \mls\ require a fairly high and stable mass-transfer rate, so their theory is not applicable to the case of \mls.

\mls\ is at least the third system in which the Stage A method yields a mass ratio in excess of $q = 0.25$, the other two being V1006~Cyg \citep{v1006cyg} and MN~Dra \citep{MN_Dra}. The fact that the Stage A method yields a mass ratio above $q = 0.25$ for these systems suggests that either the period of Stage A superhumps is not purely dynamical \citep[in contradiction of ][]{ko13} or that the disk can become eccentric at mass ratios higher than predicted by simulations \citep[in contradiction of simulations; e.g., ][]{smith07}. As our referee, Taichi Kato, pointed out to us, the first option is unlikely; the pressure effect that causes the superhump period to shorten during Stage B decreases the disk's precessional rate, so if it were present during Stage A as well, the result would be a shorter Stage A period---and, therefore, a lower mass ratio. Conversely, smoothed-particle-hydrodynamics simulations of disks generally do not model the disk-instability mechanism or the resulting changes in the disk's radius, raising the possibility that the 3:1 resonance may be achieved more easily than these simulations predict.

\subsection{Shrinkage of the disk after the superoutburst}
\label{disk_size}

One of the core predictions of the TTI model is that the radius of the accretion disk gradually increases across a supercycle, with a minimum radius after the superoutburst \citep{osaki89}. The behavior of the stream-disk hotspot in \mls\ provides evidence of this phenomenon.

\begin{figure}
\epsscale{1.2}
\plotone{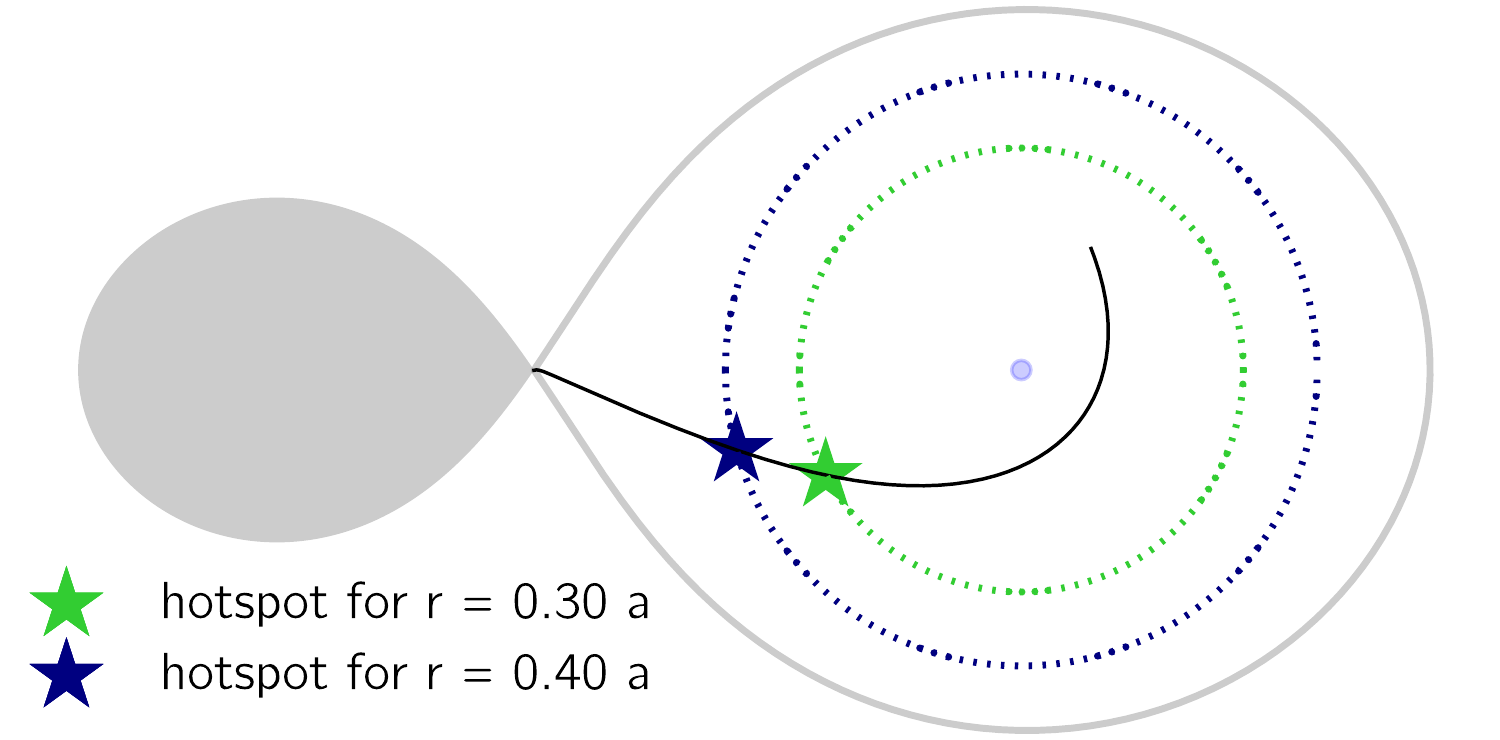}
\caption{Roche geometry for MLS, with the ballistic trajectory of the accretion stream plotted as a black line. For two different disk radii, the point of collision between the stream and the outer rim of the disk (dashed lines) is shown. \label{hotspot_schematic}}
\end{figure}

As discussed previously in Sec~\ref{hotspot_sec} and shown in Fig.~\ref{hotspot_fig}, the hotspot shifted towards earlier orbital phases after the superoutburst. We computed the coordinates within the binary rest frame of the stream-disk collision, assuming a ballistic trajectory from the L1 point for $q = 0.28$, for different values of the disk radius between $0.25a$ and $0.4a$, where $a$ is the orbital separation.\footnote{These calculations treat the hotspot as a point-source on a perfectly circular disk rim that is eclipsed by a spherical donor star.} The schematic diagram in Fig.~\ref{hotspot_schematic} shows that as the disk radius shrinks, the hotspot is viewed most directly at earlier orbital phases. For example, if the disk radius were to shrink from r = 0.4$a$ to 0.3$a$, the hotspot would be seen face-on $\sim$0.03 phase units earlier.

Using the coordinates of the stream-disk collision at different disk radii, we calculated the orbital phases at which that point would undergo eclipse ingress and egress. We find that as the disk radius expands, the eclipse of the hotspot becomes longer and ends later. Thus, a shrinkage of the disk radius after the superoutburst could account for the observed decrease of the eclipse duration. 

Moreover, the changed location of the hotspot resulted in a significant change in its appearance after the eclipse. Prior to the superoutburst, the declining edge of the hotspot was visible after the eclipse, but this feature disappeared after the superoutburst. The most straightforward interpretation is that disk radius shrank after the superoutburst, enabling the stream to travel farther along its ballistic trajectory before striking the disk. In principle, this change should also lead to an increased luminosity of the hotspot after a superoutburst, since the stream-disk collision is deeper in the WD's gravitational potential. This effect would probably be far easier to detect in a non-eclipsing system, as the presence of eclipses greatly complicates efforts to identify changes in the spot's luminosity.

\subsection{Superhump growth rate}

As stated in section~\ref{SO_sec} and shown in Fig.~\ref{superhump_fig}, the superhumps in \mls\ became apparent only $\sim$18 cycles after the start of the precursor outburst and only $\sim$9 cycles after the precursor's maximum. These short times are consistent with \citet{lubow91a, lubow91b} who calculated that growth rates to reach the 3:1 resonance should be proportional to $q^{2}$. At the low $q$ values typically observed in very short-period SU UMa systems, the appearance of superhumps takes about 60 cycles for $q = 0.06$, so the faster time for the higher $q$ of \mls\ is expected. The Stage A duration of 9-10 cycles is in accord with a high value of $q$, as shown from the compilation in \citet{kato15}.

However, the fast rise of Stage A and its short duration do not agree with the values for the high-$q$ ($\ge$0.26) system V1006 Cyg \citep{v1006cyg}, which took about 30 cycles for the superhumps to appear and in which Stage A lasted for at least 32 cycles. MN Dra also appears to have a large $q$ value (0.29) and a long Stage A \citep{MN_Dra}. \citet{v1006cyg} postulated that systems with mass ratios near the tidal-stability limit might have slow superhump growth rates. Both the MN Dra and V1006 Cyg datasets suffered from lack of data at the start of the outburst, and both systems have somewhat peculiar orbital periods in the period gap, so it will require further data on high-$q$ systems to determine how superhump development is affected by a high mass ratio.

\subsection{Minioutbursts}
\label{mini}

During the best-defined minioutburst, the eclipse depth increased, as did the FWHM, consistent with the extra luminosity originating in the outer disk. If the minioutbursts originated in the inner disk or on the WD, the eclipses would have immediately become deeper and narrower. The minioutbursts ceased in the wake of the superoutburst, implying that their appearance was closely linked to disk changes during the supercycle or to accretion episodes that ended after the superoutburst.

\citet{ok14} reported the detection of minioutbursts in Kepler observations of V1504 Cyg and proposed that they were related to the increased disk radius during the supercycle. Specifically, they argued that tidal dissipation in the outer disk could prematurely trigger a thermal instability, causing a brief outburst with a diminished amplitude. In their explanation, the disk then jumps from the cold branch of the thermal equilibrium curve to an intermediate warm branch (as opposed to the hot branch, as would occur during a normal outburst). Because the TTI model predicts that the disk radius will increase during the supercycle, it offers a plausible explanation as to why the minioutbursts in \mls\ are only observed before the superoutburst. A theoretical examination of this proposal would be a logical next step.

It is also possible that a beat phenomenon between the orbital signal and failed superhumps could produce the minioutbursts. Failed superhumps are observed during the declining portion of normal outbursts before a superoutburst and fall into two general categories---positive and negative---depending, respectively, on whether their period is longer \citep{om03} or shorter \citep[][ and references therein]{ok13} than the orbital period.

Because the maxima of the minioutbursts occur quasi-periodically every $\sim$2~d, the 1.909-h orbital period would need to beat against a periodicity of roughly $\sim$1.84~h (0.545 cycles hr$^{-1}$) or $\sim$1.99~h (0.503 cycles hr$^{-1}$), since $\nu_{beat} = | \nu_{orbit} - \nu_{SH} | $. In the trailed power spectrum in Fig.~\ref{full_LC}, there is a brief ($\sim$0.5~d) oscillation near the peak of the second normal outburst, during which the power shifts from the orbital period to a period of $\sim$1.96~h. As this is longer than the orbital period, this could be attributable to failed positive superhumps, which are thought to arise when a normal outburst is extinguished before the tidal instability has had enough time to fully develop in disk material at the 3:1 resonance radius \citep{om03}. According to \citet{om03}, this underdeveloped tidal instability should not persist into quiescence in a high-mass-ratio system, making it difficult to envision how it could account for minioutbursts during quiescence. Moreover, a 1.96-h period would produce a beat period of $\sim$3~d with the orbital period. Even taking into account the quasi-periodic nature of the minioutbursts, this period is longer than the typical interval between them, making it unlikely that the $\sim$1.96-hr period is associated with the minioutbursts. 

While our power spectra do not show evidence of a superhump signal that could produce the minioutbursts by beating against the orbital period, the orbital modulation is so strong that it could potentially obscure the presence of a transitory, low-amplitude superhump signal. Thus, we cannot entirely rule out this possibility.

\section{Conclusion}

\mls\ is the first eclipsing SU UMa-type system for which a superoutburst has been observed by \textit{Kepler} in the short-cadence mode. There were eight normal, outside-in outbursts, one of which was a precursor to the superoutburst. Superhumps emerged near the maximum of the precursor and reached their maximum amplitude of $\sim$0.4~mag in just several orbits. The superhump amplitude fluctuated during the early-to-mid superoutburst and appeared to correlate with the orbital phase at which the superhump maximum occurred. This effect could be caused by orbital-phase-dependent obscuration of the superhump light source by an elevated, non-axisymmetric disk rim, or it could be related to the viewing aspect of the intrinsically asymmetric superhump light source as suggested by \citet{om03}.

The mass ratio of \mls, estimated to be $q = 0.281\pm0.015$ from the period excess of Stage A superhumps, is marginally inconsistent with simulations of superhumps that predict a limiting mass ratio of $q = 0.25$ for superhump formation. However, the overlap between eclipses and half of the Stage A superhump maxima means that the uncertainty of our measurement might be underestimated, so an independent measurement of the mass ratio in a follow-up study would be very useful.

We detected a phase shift of the stream-disk hotspot towards earlier orbital phases after the superoutburst. We attribute the shift to a shrinkage of the disk radius after the superoutburst, as predicted by the TTI model.

\mls\ also displayed a series of unusual minioutbursts that abruptly ceased after the superoutburst. Their cause remains elusive, and it would be beneficial if a future theoretical study were to attempt to incorporate them into the TTI model.

\acknowledgments

We thank the referee, Taichi Kato, for a comprehensive and expeditious report that significantly improved this paper.

P.S. acknowledges support from NSF grant AST-1514737 and is grateful for discussions of this object at the KITP Conference on Confronting MHD Theories of Accretion Disks with Observations. 

Z.D. is supported by CAS Light of West China Program and the Science Foundation of Yunnan Province (No. 2016FB007). We thank Dr. Kai Li for observing \mls\ at Weihai Observatory.

\end{document}